\documentclass[
 aip,
 amsmath,amssymb,
 reprint,
]{revtex4-1}

\usepackage{chemformula}
\usepackage{bm} 
\usepackage[utf8]{inputenc}
\usepackage[T1]{fontenc}
\usepackage{mathptmx}
\usepackage{etoolbox}
\usepackage{subfiles}

\makeatletter
\def\@email#1#2{
 \endgroup
 \patchcmd{\titleblock@produce}
  {\frontmatter@RRAPformat}
  {\frontmatter@RRAPformat{\produce@RRAP{*#1\href{mailto:#2}{#2}}}\frontmatter@RRAPformat}
  {}{}
}
\makeatother
\begin{document}

\preprint{AIP/123-QED}

\title{Comparative analysis of spin wave imaging using nitrogen vacancy centers and time resolved magneto-optical measurements}

\author{Carolina Lüthi}
\affiliation{Physics Department, Technical University of Munich, Garching, Germany}

\author{Lukas Colombo}
\affiliation{Physics Department, Technical University of Munich, Garching, Germany}

\author{Franz Vilsmeier}
\affiliation{Physics Department, Technical University of Munich, Garching, Germany}

\author{Christian Back}
\affiliation{Physics Department, Technical University of Munich, Garching, Germany}
\affiliation{Munich Center for Quantum Science and Technology (MCQST), Munich, Germany}
\email{christian.back@tum.de}

\date{\today}

\begin{abstract}
Spin waves, the fundamental excitations in magnetic materials, are promising candidates for realizing low-dissipation information processing in spintronics. The ability to visualize and manipulate coherent spin-wave transport is crucial for the development of spin wave-based devices. We use a recently discovered method utilizing nitrogen vacancy (NV) centers, point defects in the diamond lattice, to measure spin waves in thin film magnetic insulators by detecting their magnetic stray field. We experimentally demonstrate enhanced contrast in the detected wavefront amplitudes by imaging spin waves underneath a reference stripline and phenomenologically model the results. By extracting the spin wave dispersion and comparing NV center based spin wave measurements to spin wave imaging conducted through the well-established time-resolved magneto-optical Kerr effect, we discuss the advantages and limitations of employing NV centers as spin wave sensors.
\end{abstract}

\maketitle

Spin waves\cite{Bloch.1930, Gurevich.1996, Stancil.2008}, also known as magnons, represent collective excitations of the spins in a magnetic material. These quasi-particles carry spin angular momentum and propagate through the material's magnetic lattice, thereby avoiding heating associated with charge currents. Their unique properties, including long coherence times and low dissipation \cite{Wang.2023, Kruglyak.2010, Hortensius.2021, Serga.2010, Chumak.2014}, as well as frequencies in the giga- to terrahertz regime and wavelength as small as several nanometers \cite{Cherepanov.1993, Balashov.2014, Chuang.2014}, render spin waves promising candidates for various applications, such as interference-based ultrafast, nanoscale magnonic logic circuits, and miniaturized device technologies \cite{Hortensius.2021, Chumak.2015, Hahn.2014, Chumak.2014, Khitun.2012, Khitun.2010, Schneider.2008b, Owens.1985}, particularly in the field of spintronics\cite{Han.2023, Singh.2023, Zhang.2023, Chumak.2015}. 

Imaging spin waves with high spatial and temporal resolution is essential for exploiting their potential in technological advancements and understanding their intricate dynamics. Traditional imaging techniques, such as time-resolved magneto-optical Kerr effect (TR-MOKE) \cite{Vilsmeier.23.03.2024, Perzlmaier.2008, Farle.2013, Au.2011, Bauer.2014, Stigloher.2018}, Brillouin light scattering \cite{Hillebrands.1999, Demokritov.2001}, and transmission x-ray microscopy \cite{Warwick.1998, Sluka.2019}, have provided valuable insights into spin wave behavior. However, these methods often have limitations in terms of spatial resolution and sensitivity, particularly when studying spin waves in nanoscale systems or beneath opaque materials, or, in the case of x-ray microscopy, need large and expensive facilities for their implementation.

In recent years, the emergence of nitrogen-vacancy (NV) centers in diamond as sensitive magnetic field sensors has opened new avenues for spin wave imaging \cite{Bertelli.2020, Purser.2020, Koerner.2022, Tetienne.2013}. NV centers, point-like defects in the diamond lattice consisting of a substitutional nitrogen atom adjacent to a vacancy, exhibit remarkable properties, including high sensitivity to static and fluctuating magnetic fields \cite{Maletinsky.2012, Appel.2015, Rondin.2014}, nanoscale spatial resolution \cite{Balasubramanian.2008, vanderSar.2015}, even below opaque materials\cite{Bertelli.2021}, and optical addressability \cite{Doherty.2013}. By using an ensemble of spins instead of a single NV center, the sensitivity can be further enhanced by a scaling of $\sqrt{N}$, where $N$ is the number of NV centers\cite{Purser.2020}.

In this work, we discuss the use of an ensemble of shallowly implanted NV centers as a means to measure spin waves in a magnetic thin film following the seminal work reported in \cite{vanderSar.2015}, in this case the low damping ferrimagnetic insulator yttrium iron garnet (YIG) \cite{Cherepanov.1993, Serga.2010}. We demonstrate how the contrast in the imaged spin wave amplitude can be enhanced, allowing us to image spin waves for a long travelling distance. Furthermore, we phenomenologically model these measurements to explain the observed features in the NV spin wave images. By comparing the spin wave measurements obtained through NV center detection to spin wave imaging acquired with conventional TR-MOKE measurements, we discuss the advantages and limitations of utilizing NV centers for spin wave imaging, as well as the potential of integrating NV center measurements with TR-MOKE imaging into a unified setup.

\begin{figure}[htb!]
    \centering
    \includegraphics[width=0.48\textwidth]{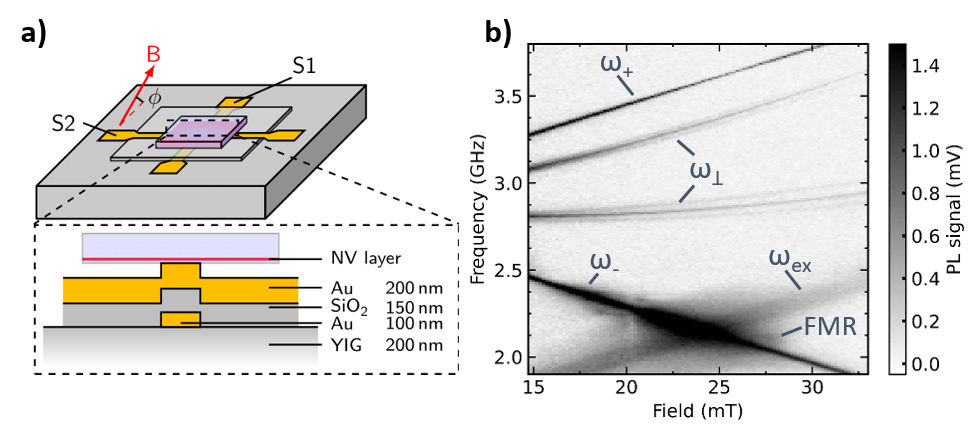}
    \caption{a) Sample configuration: A diamond chip containing a layer of NVs is positioned on a 200~nm thick YIG film grown on a gadolinium gallium garnet substrate. A stripline S1 excites spin waves in the sample. A second perpendicular stripline S2, separated from S1 by a $\mathrm{SiO_2}$ layer, is used to generate an Oerstedfield $B_{\mathrm{oe}}$ that oscillates at the same frequency as the spin wave strayfield. An external magnetic field $B$ is applied along S1 at an angle of $\phi = 35^\circ$ relative to the sample plane. b) Photoluminescence spectrum of the NV centers in dependence of the external magnetic field $B$ and the drive frequency. Transitions in the ground ($\omega_\pm,\omega_\perp$) and excited states ($\omega_\mathrm{ex}$) correspond to a peak in the signal. Due to the handedness of the circularly polarized spin wave stray field, the $\omega_-$ transitions are enhanced for frequencies above the ferromagnetic resonance (FMR) limit.}
    \label{fig:sample}
\end{figure}

\begin{figure*}[tb!]
    \centering
    \includegraphics[width=0.85\textwidth]{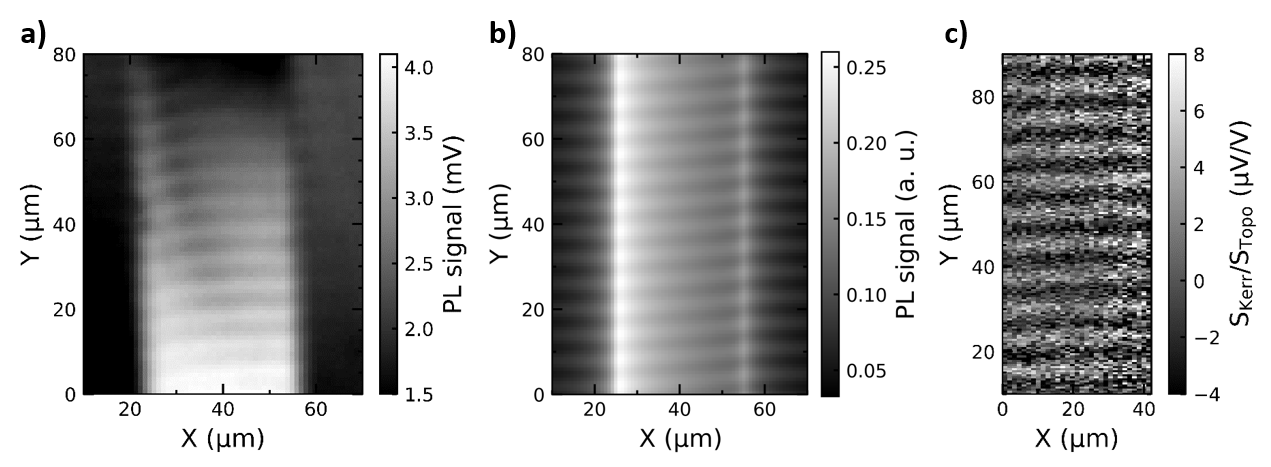}
    \caption{a) Spatially resolved PL signal measured above the YIG film, when a spin wave with frequency $f_\mathrm{sw} = 2.24$~GHz, corresponding to the external field $B = 22.5$~mT, is excited by S1. The X-axis is located 10~$\mathrm{\mu}$m away from the center of S1, and the center of S2 is located at $X = 40~\mathrm{\mu m}$. $B$ is applied along S1 at an angle of $\phi = 35^\circ$ relative to the sample plane. The microwave current is split between S1 and S2 to additionally generate an Oersted field $B_\mathrm{oe}$ oscillating at $f_\mathrm{sw}$, which results, together with the strayfield of the spin wave, in a standing wave of the total magnetic field above the YIG film. b) Simulated PL signal corresponding to the same spatial position as in a). The phase shift of the simulated wavefronts at the edges of S2 occur due to the change in sign of $B_\mathrm{oe}$ from the left side to the right side of S2. c) Normalized TR-MOKE signal, when a spin wave with frequency $f_\mathrm{sw} = 2.24$~GHz is excited by S1. The external field is applied along S1, which is approximately 10~$\mathrm{\mu}$m away from the X-axis.}
    \label{fig:sw}
\end{figure*}

We excite spin waves in a 200~nm thick YIG thin film grown by liquid phase epitaxy on a gadolinium gallium garnet substrate by a microwave current sent through a stripline (S1) fabricated onto the YIG surface. The diamond chip is placed on the YIG sample with the NV layer facing towards the magnetic surface, as shown in Fig.~\ref{fig:sample}.~a). Additional information on the diamond chip and the sample can be found in the Supplementary Material \ref{A:sample}. As the spin waves propagate right (left)-wards, they generate a circularly polarized field with handedness that drives the $\omega_-  (\omega_+)$ ESR transition of the NVs, resulting in a spin-dependent photoluminescence (PL) signal\cite{Bertelli.2020, vanderSar.2015, Andrich.2017}.
In Fig.~\ref{fig:sample}.~b) the NV photoluminescence signal, obtained via lock-in detection and measured approximately 10~$\mathrm{\mu}$m away from the center of S1, is shown in dependence of the external magnetic field $B$ and the microwave current frequency. The NV setup is described in detail in the Supplementary Material \ref{A:setup}. Here, $B$ is applied along S1 at an angle of $\phi = 35^\circ$ relative to the sample plane, aligning it with one of the four possible orientations of the NV axes. The peaks in the PL signal belong to the ESR transitions in the ground ($\omega_\pm,\omega_\perp$) and excited states ($\omega_\mathrm{ex}$), where only the PL signal of NV centers aligned with $B$ result in a linear dependence on $B$. As the NVs are not only driven by the microwave field of S1 but additionally also by $B_{\mathrm{sw}}$, the $\omega_-$ transitions are enhanced for frequencies above the ferromagnetic resonance (FMR) limit.

To gain the phase sensitivity to image individual wave fronts of the spin waves, a second perpendicular reference stripline (S2), isolated from S1 by an insulating $\mathrm{SiO_2}$ layer, is used (Fig.~\ref{fig:sample}.~a)). A microwave current that generates an Oersted field $B_{\mathrm{oe}}$ oscillating at the same frequency as $B_{\mathrm{sw}}$ is sent through S2.
The superposition of $B_{\mathrm{oe}}$ with the spin wave stray field $B_{\mathrm{sw}}$ results in a standing wave pattern in the total magnetic field that drives the NV ESR with a spatial periodicity equal to the spin wave wavelength \cite{Bertelli.2020}. In Fig.~\ref{fig:sw}.~a), the PL signal for a spatial scan over the sample is shown. The X-axis, aligned with S1, is located approximately 10~$\mathrm{\mu}$m away from the center of S1. Furthermore, we scanned over S2, which is aligned with the Y-axis, with its center located at $X = 40~\mathrm{\mu m}$. We find an enhanced contrast in the imaged spin wave amplitude directly above S2. Additionally, we observe a $180^\circ$ phase shift of the wave fronts at the edges of S2 and a light bending of the wave fronts towards the center of S2. 

To explain these characteristics, we phenomenologically model our PL signal by calculating the time-averaged magnitude of the perpendicular time dependent magnetic field $B_\mathrm{tot_\perp}$ relative to the NV axis. Thereby, we take into account that ESR transitions can only be driven by rf fields perpendicular to their quantization axis. Furthermore, we average over time, as we measure a time independent PL signal at a fixed position on the sample, which scales with the magnitude of the standing wave in the total perpendicular magnetic field. We use a coordinate system where the X-axis is aligned with S1, the Y-axis with S2, and the Z-axis with the normal vector of the sample plane. In $B_\mathrm{tot_\perp}$, we consider the spin wave strayfield above the film $B_\mathrm{sw}$ as well as the Oersted field $B_\mathrm{oe}$ of S2:
\begin{equation}
    B_\mathrm{tot_\perp}(t)  = B_\mathrm{sw}(t) + B_\mathrm{oe}(t).
\end{equation}
The time dependent magnetic fields $B_\mathrm{sw}(t)$ and $B_\mathrm{oe}(t)$, as well as further details about the simulation, are given in the Supplementary Material \ref{A:simulation}.
In Fig.~\ref{fig:sw}.~b), the simulated PL signal corresponding to the same spatial position on the sample as the measurement depicted in Fig.~\ref{fig:sw}.~a) is shown. The observed phase shift at the edges of S2 is attributed to the change in sign of $B_\mathrm{oe}$ from the left side to the right side of S2. Moreover, we find an enhanced contrast in the PL signal above S2 due to the comparable magnitudes of $B_\mathrm{sw}(t)$ and $B_\mathrm{eo}(t)$ in the vicinity of S2.  However, reproducing the bending of the wavefronts towards the center of S2 requires a more advanced model, which is beyond the scope of this study.

In the following, we compare our NV measurements to TR-MOKE measurements in Damon-Eshbach (DE) configuration\cite{Damon.1961}, where the external field is applied in the YIG thin film plane along S1. TR-MOKE measurements involve synchronizing the rf excitation and optical probing pulses. Similar to stroboscopic techniques, a time-independent phase relationship between excitation and probing is typically established to capture a stationary image. To achieve this synchronization, a pulsed Ti:Sa-laser with a repetition rate $f_\mathrm{rep} = 80$~MHz is employed, while the excitation frequency $f_\mathrm{ex}$ of the spin waves fulfils $f_\mathrm{ex} = n \cdot f_\mathrm{rep}$, with $n$ being an integer. On the other hand, to detect spin waves with frequencies unequal to multiples of $f_\mathrm{rep}$, we use super-Nyquist sampling MOKE \cite{Dreyer.2021}. In that case, we can tune the rf excitation to any intermediate frequency $f_\mathrm{ex} = n \cdot f_\mathrm{rep} + \epsilon$, where $\epsilon$ is a rational number. Demodulating the Kerr signal at frequency $\epsilon$ enables direct extraction of the real and imaginary components of the magnetic rf-susceptibility, thereby providing phase-resolved measurements of the spin precession \cite{Dreyer.2021} (See Supplementary Material \ref{A:setup} for further information concerning the MOKE setup). 
The lock-in detected Kerr signal $S_\mathrm{Kerr}$ is then normalized to the topographic signal $S_\mathrm{Topo}$ to gain an enhanced contrast in the measured spin wavefronts. Fig. ~\ref{fig:sw}.~c) shows a TR-MOKE measurement of propagating spin waves. As direct contact with the magnetic surface is needed, the measurement is taken far away from S2. In Supplementary Material \ref{A:2D wavefronts}, we additionally show spin wave measurements with different wavelengths.

\begin{figure}[htb!]
    \centering
    \includegraphics[width=0.3\textwidth]{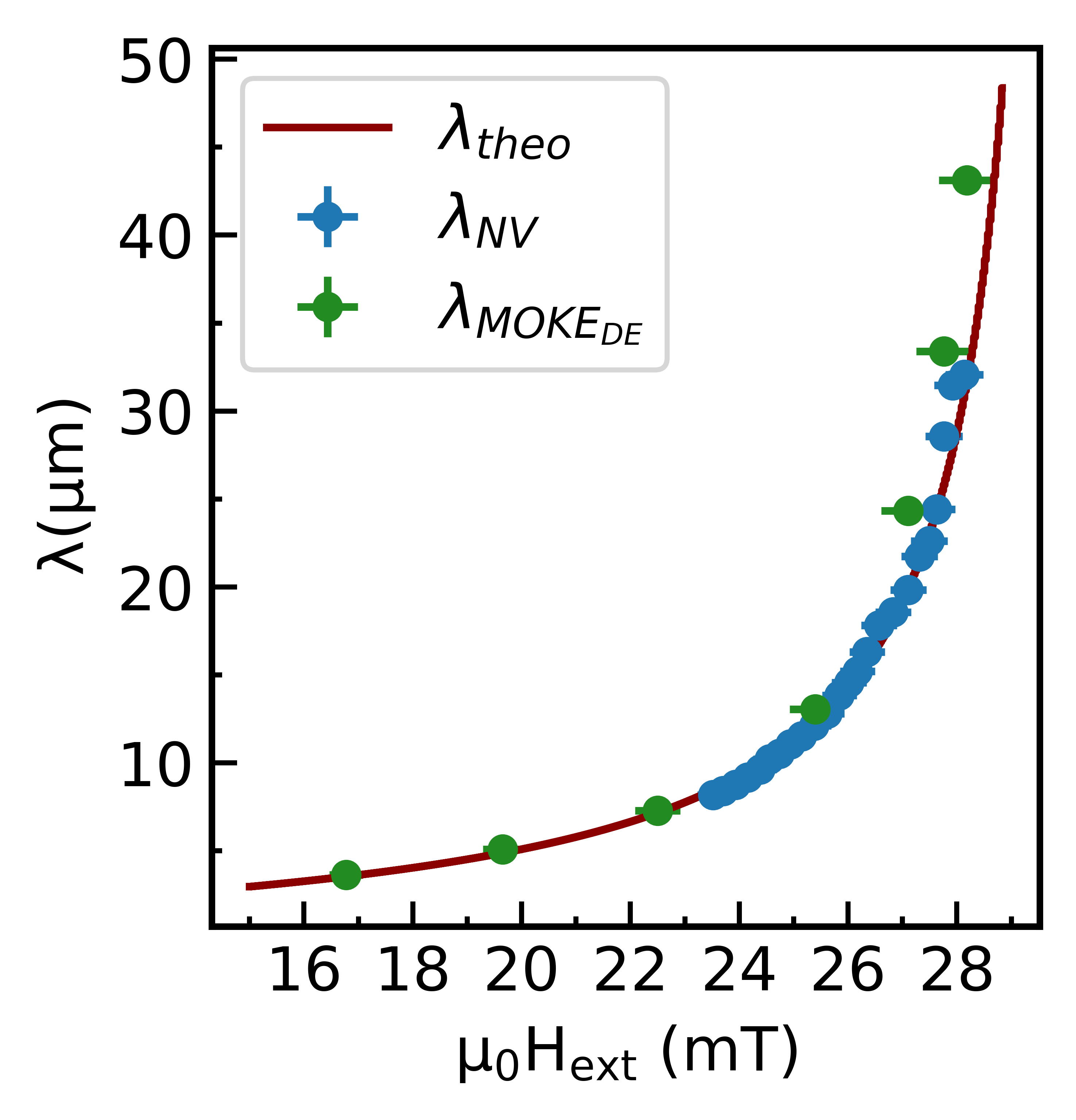}
    \caption{Dependence of the wavelength of the excited spin waves on the external magnetic field. The theoretical dispersion curve (red line) is in good agreement with the fitted wavelength from the NV measurements (blue dots) and the TR-MOKE measurements (green dots). }
    \label{fig:dispersion}
\end{figure}

To extract the spin wave dispersion from both measurement techniques, we adjust the external static magnetic field $B$ such that the spin wave frequency coincides with the $\omega_-$ NV ESR frequencies (The mathematical equations to calculate the dispersion are given in Supplementary Material \ref{A:Dispersion}). This allows us to probe spin waves with varying wavelength, detectable by both measurement techniques. In the NV measurements, the field is applied along S1 and one of the NV quantization axes. Given that the external field is always less than 30~mT, significantly smaller than the saturation magnetization of YIG $\mu_0 M_\mathrm{S} \thickapprox 185$~mT, the static magnetization of the YIG film tilts only slightly out of plane, at an angle $B \mathrm{sin}(\phi) / (\mu_0 M_\mathrm{S}) \leq 5.3^\circ.$ Hence, direct comparison between the NV measurements and the in-plane MOKE measurements is feasible. Additionally, in the Supplementary Material \ref{A:Moke DE vs NV}, we demonstrate that within this range of magnetic fields, MOKE measurements in DE configuration and with the field applied along the NV axis yield consistent wavelengths within the error margin. To  facilitate a direct comparison of the measured wavelengths, we transpose the in plane fields of the TR-MOKE measurements into the direction of the NV center axis. 
In Fig.~\ref{fig:dispersion}, the theoretical dispersion curve (red line), alongside the fitted wavelength of the NV measurement (blue dots) and the MOKE measurements (green dots) are shown, which are in good agreement with each other (See supplementary material \ref{A:fit model} for information about the fit model.). This consistency demonstrates the reliability and equivalence of the two measurement techniques.

In conclusion, the utilization of NV centers presents several advantages in the measurement of spin waves. Their high resolution capabilities extend beyond the diffraction limit, enabling precise characterization of spin wave phenomena. Moreover, their ability to measure through opaque materials offers unique opportunities for non-invasive investigation \cite{Bertelli.2021, Borst.2023}. Additionally, NV centers exhibit remarkable sensitivity to magnetic fields, enhancing their utility in detecting subtle magnetic variations.
However, it is worth noting some limitations of NV centers, including the lack of time resolution and the requirement of a standing wave for effectively imaging wavefronts. Additionally, NV centers are limited to probing only spin waves at the NV ESR frequencies.
On the other hand, Kerr microscopy offers high time resolution and the flexibility to probe a wide range of frequencies. However, direct contact with the magnetic surface is necessary, and the  spacial resolution is ultimately limited by diffraction.
To overcome these limitations and harness the complementary strengths of both techniques, we propose a combined approach. By integrating NV center measurements with Kerr microscopy, one can benefit from high-resolution imaging, non-invasive probing through opaque materials, high sensitivity to magnetic fields, high time resolution, and flexibility in probing frequencies. This integrated approach promises to advance the capability to probe spin wave dynamics and enable new insights in the field.

\begin{acknowledgments}
	This work was supported by the Bayerisches Staatsministerium f\"ur
Wissenschaft und Kunst through project IQSense via the Munich Quantum Valley (MQV) and by the DFG via the Munich Center for Quantum Science and Technology (MCQST, EXC2111).
 
\end{acknowledgments}

\section*{Data Availability Statement}
The data that support the findings of this study are available from the corresponding author upon reasonable request.

\nocite{*}
\bibliography{bib}

\begin{thebibliography}{54}%
\makeatletter
\providecommand \@ifxundefined [1]{%
 \@ifx{#1\undefined}
}%
\providecommand \@ifnum [1]{%
 \ifnum #1\expandafter \@firstoftwo
 \else \expandafter \@secondoftwo
 \fi
}%
\providecommand \@ifx [1]{%
 \ifx #1\expandafter \@firstoftwo
 \else \expandafter \@secondoftwo
 \fi
}%
\providecommand \natexlab [1]{#1}%
\providecommand \enquote  [1]{``#1''}%
\providecommand \bibnamefont  [1]{#1}%
\providecommand \bibfnamefont [1]{#1}%
\providecommand \citenamefont [1]{#1}%
\providecommand \href@noop [0]{\@secondoftwo}%
\providecommand \href [0]{\begingroup \@sanitize@url \@href}%
\providecommand \@href[1]{\@@startlink{#1}\@@href}%
\providecommand \@@href[1]{\endgroup#1\@@endlink}%
\providecommand \@sanitize@url [0]{\catcode `\\12\catcode `\$12\catcode
  `\&12\catcode `\#12\catcode `\^12\catcode `\_12\catcode `\%12\relax}%
\providecommand \@@startlink[1]{}%
\providecommand \@@endlink[0]{}%
\providecommand \url  [0]{\begingroup\@sanitize@url \@url }%
\providecommand \@url [1]{\endgroup\@href {#1}{\urlprefix }}%
\providecommand \urlprefix  [0]{URL }%
\providecommand \Eprint [0]{\href }%
\providecommand \doibase [0]{http://dx.doi.org/}%
\providecommand \selectlanguage [0]{\@gobble}%
\providecommand \bibinfo  [0]{\@secondoftwo}%
\providecommand \bibfield  [0]{\@secondoftwo}%
\providecommand \translation [1]{[#1]}%
\providecommand \BibitemOpen [0]{}%
\providecommand \bibitemStop [0]{}%
\providecommand \bibitemNoStop [0]{.\EOS\space}%
\providecommand \EOS [0]{\spacefactor3000\relax}%
\providecommand \BibitemShut  [1]{\csname bibitem#1\endcsname}%
\let\auto@bib@innerbib\@empty
\bibitem [{\citenamefont {Bloch}(1930)}]{Bloch.1930}%
  \BibitemOpen
  \bibfield  {author} {\bibinfo {author} {\bibfnamefont {F.}~\bibnamefont
  {Bloch}},\ }\bibfield  {title} {\enquote {\bibinfo {title} {Zur {T}heorie des
  {F}erromagnetismus},}\ }\href {\doibase 10.1007/BF01339661} {\bibfield
  {journal} {\bibinfo  {journal} {Zeitschrift f{\"u}r Physik}\ }\textbf
  {\bibinfo {volume} {61}},\ \bibinfo {pages} {206--219} (\bibinfo {year}
  {1930})}\BibitemShut {NoStop}%
\bibitem [{\citenamefont {Gurevich}\ and\ \citenamefont
  {Melkov}(1996{\natexlab{a}})}]{Gurevich.1996}%
  \BibitemOpen
  \bibfield  {author} {\bibinfo {author} {\bibfnamefont {A.~G.}\ \bibnamefont
  {Gurevich}}\ and\ \bibinfo {author} {\bibfnamefont {G.~A.}\ \bibnamefont
  {Melkov}},\ }\href {\doibase 10.1201/9780138748487} {\emph {\bibinfo {title}
  {Magnetization oscillations and waves}}}\ (\bibinfo  {publisher} {{CRC
  Press}},\ \bibinfo {address} {Boca Raton, Florida},\ \bibinfo {year}
  {1996})\BibitemShut {NoStop}%
\bibitem [{\citenamefont {Stancil}\ and\ \citenamefont
  {Prabhakar}(2008)}]{Stancil.2008}%
  \BibitemOpen
  \bibfield  {author} {\bibinfo {author} {\bibfnamefont {D.~D.}\ \bibnamefont
  {Stancil}}\ and\ \bibinfo {author} {\bibfnamefont {A.}~\bibnamefont
  {Prabhakar}},\ }\href {\doibase 10.1007/978-0-387-77865-5} {\emph {\bibinfo
  {title} {Spin Waves: Theory and Applications}}},\ \bibinfo {edition} {1st}\
  ed.\ (\bibinfo  {publisher} {{Springer US}},\ \bibinfo {address} {Berlin},\
  \bibinfo {year} {2008})\BibitemShut {NoStop}%
\bibitem [{\citenamefont {Wang}\ \emph {et~al.}(2023)\citenamefont {Wang},
  \citenamefont {Yuan}, \citenamefont {Zhou}, \citenamefont {Zhang},
  \citenamefont {Chen}, \citenamefont {Liu}, \citenamefont {Jia}, \citenamefont
  {Yu}, \citenamefont {Ansermet}, \citenamefont {Song},\ and\ \citenamefont
  {Yu}}]{Wang.2023}%
  \BibitemOpen
  \bibfield  {author} {\bibinfo {author} {\bibfnamefont {H.}~\bibnamefont
  {Wang}}, \bibinfo {author} {\bibfnamefont {R.}~\bibnamefont {Yuan}}, \bibinfo
  {author} {\bibfnamefont {Y.}~\bibnamefont {Zhou}}, \bibinfo {author}
  {\bibfnamefont {Y.}~\bibnamefont {Zhang}}, \bibinfo {author} {\bibfnamefont
  {J.}~\bibnamefont {Chen}}, \bibinfo {author} {\bibfnamefont {S.}~\bibnamefont
  {Liu}}, \bibinfo {author} {\bibfnamefont {H.}~\bibnamefont {Jia}}, \bibinfo
  {author} {\bibfnamefont {D.}~\bibnamefont {Yu}}, \bibinfo {author}
  {\bibfnamefont {J.-P.}\ \bibnamefont {Ansermet}}, \bibinfo {author}
  {\bibfnamefont {C.}~\bibnamefont {Song}}, \ and\ \bibinfo {author}
  {\bibfnamefont {H.}~\bibnamefont {Yu}},\ }\bibfield  {title} {\enquote
  {\bibinfo {title} {Long-distance coherent propagation of high-velocity
  antiferromagnetic spin waves},}\ }\href {\doibase
  10.1103/PhysRevLett.130.096701} {\bibfield  {journal} {\bibinfo  {journal}
  {Physical review letters}\ }\textbf {\bibinfo {volume} {130}},\ \bibinfo
  {pages} {096701} (\bibinfo {year} {2023})}\BibitemShut {NoStop}%
\bibitem [{\citenamefont {Kruglyak}, \citenamefont {Demokritov},\ and\
  \citenamefont {Grundler}(2010)}]{Kruglyak.2010}%
  \BibitemOpen
  \bibfield  {author} {\bibinfo {author} {\bibfnamefont {V.~V.}\ \bibnamefont
  {Kruglyak}}, \bibinfo {author} {\bibfnamefont {S.~O.}\ \bibnamefont
  {Demokritov}}, \ and\ \bibinfo {author} {\bibfnamefont {D.}~\bibnamefont
  {Grundler}},\ }\bibfield  {title} {\enquote {\bibinfo {title} {Magnonics},}\
  }\href {\doibase 10.1088/0022-3727/43/26/264001} {\bibfield  {journal}
  {\bibinfo  {journal} {Journal of Physics D: Applied Physics}\ }\textbf
  {\bibinfo {volume} {43}},\ \bibinfo {pages} {264001} (\bibinfo {year}
  {2010})}\BibitemShut {NoStop}%
\bibitem [{\citenamefont {Hortensius}\ \emph {et~al.}(2021)\citenamefont
  {Hortensius}, \citenamefont {Afanasiev}, \citenamefont {Matthiesen},
  \citenamefont {Leenders}, \citenamefont {Citro}, \citenamefont {Kimel},
  \citenamefont {Mikhaylovskiy}, \citenamefont {Ivanov},\ and\ \citenamefont
  {Caviglia}}]{Hortensius.2021}%
  \BibitemOpen
  \bibfield  {author} {\bibinfo {author} {\bibfnamefont {J.~R.}\ \bibnamefont
  {Hortensius}}, \bibinfo {author} {\bibfnamefont {D.}~\bibnamefont
  {Afanasiev}}, \bibinfo {author} {\bibfnamefont {M.}~\bibnamefont
  {Matthiesen}}, \bibinfo {author} {\bibfnamefont {R.}~\bibnamefont
  {Leenders}}, \bibinfo {author} {\bibfnamefont {R.}~\bibnamefont {Citro}},
  \bibinfo {author} {\bibfnamefont {A.~V.}\ \bibnamefont {Kimel}}, \bibinfo
  {author} {\bibfnamefont {R.~V.}\ \bibnamefont {Mikhaylovskiy}}, \bibinfo
  {author} {\bibfnamefont {B.~A.}\ \bibnamefont {Ivanov}}, \ and\ \bibinfo
  {author} {\bibfnamefont {A.~D.}\ \bibnamefont {Caviglia}},\ }\bibfield
  {title} {\enquote {\bibinfo {title} {Coherent spin-wave transport in an
  antiferromagnet},}\ }\href {\doibase 10.1038/s41567-021-01290-4} {\bibfield
  {journal} {\bibinfo  {journal} {Nature physics}\ }\textbf {\bibinfo {volume}
  {17}},\ \bibinfo {pages} {1001--1006} (\bibinfo {year} {2021})}\BibitemShut
  {NoStop}%
\bibitem [{\citenamefont {Serga}, \citenamefont {Chumak},\ and\ \citenamefont
  {Hillebrands}(2010)}]{Serga.2010}%
  \BibitemOpen
  \bibfield  {author} {\bibinfo {author} {\bibfnamefont {A.~A.}\ \bibnamefont
  {Serga}}, \bibinfo {author} {\bibfnamefont {A.~V.}\ \bibnamefont {Chumak}}, \
  and\ \bibinfo {author} {\bibfnamefont {B.}~\bibnamefont {Hillebrands}},\
  }\bibfield  {title} {\enquote {\bibinfo {title} {Yig magnonics},}\ }\href
  {\doibase 10.1088/0022-3727/43/26/264002} {\bibfield  {journal} {\bibinfo
  {journal} {Journal of Physics D: Applied Physics}\ }\textbf {\bibinfo
  {volume} {43}},\ \bibinfo {pages} {264002} (\bibinfo {year}
  {2010})}\BibitemShut {NoStop}%
\bibitem [{\citenamefont {Chumak}, \citenamefont {Serga},\ and\ \citenamefont
  {Hillebrands}(2014)}]{Chumak.2014}%
  \BibitemOpen
  \bibfield  {author} {\bibinfo {author} {\bibfnamefont {A.~V.}\ \bibnamefont
  {Chumak}}, \bibinfo {author} {\bibfnamefont {A.~A.}\ \bibnamefont {Serga}}, \
  and\ \bibinfo {author} {\bibfnamefont {B.}~\bibnamefont {Hillebrands}},\
  }\bibfield  {title} {\enquote {\bibinfo {title} {Magnon transistor for
  all-magnon data processing},}\ }\href {\doibase 10.1038/ncomms5700}
  {\bibfield  {journal} {\bibinfo  {journal} {Nature communications}\ }\textbf
  {\bibinfo {volume} {5}},\ \bibinfo {pages} {4700} (\bibinfo {year}
  {2014})}\BibitemShut {NoStop}%
\bibitem [{\citenamefont {Cherepanov}, \citenamefont {Kolokolov},\ and\
  \citenamefont {L'vov}(1993)}]{Cherepanov.1993}%
  \BibitemOpen
  \bibfield  {author} {\bibinfo {author} {\bibfnamefont {V.}~\bibnamefont
  {Cherepanov}}, \bibinfo {author} {\bibfnamefont {I.}~\bibnamefont
  {Kolokolov}}, \ and\ \bibinfo {author} {\bibfnamefont {V.}~\bibnamefont
  {L'vov}},\ }\bibfield  {title} {\enquote {\bibinfo {title} {The saga of yig:
  Spectra, thermodynamics, interaction and relaxation of magnons in a complex
  magnet},}\ }\href {\doibase 10.1016/0370-1573(93)90107-O} {\bibfield
  {journal} {\bibinfo  {journal} {Physics Reports}\ }\textbf {\bibinfo {volume}
  {229}},\ \bibinfo {pages} {81--144} (\bibinfo {year} {1993})}\BibitemShut
  {NoStop}%
\bibitem [{\citenamefont {Balashov}\ \emph {et~al.}(2014)\citenamefont
  {Balashov}, \citenamefont {Buczek}, \citenamefont {Sandratskii},
  \citenamefont {Ernst},\ and\ \citenamefont {Wulfhekel}}]{Balashov.2014}%
  \BibitemOpen
  \bibfield  {author} {\bibinfo {author} {\bibfnamefont {T.}~\bibnamefont
  {Balashov}}, \bibinfo {author} {\bibfnamefont {P.}~\bibnamefont {Buczek}},
  \bibinfo {author} {\bibfnamefont {L.}~\bibnamefont {Sandratskii}}, \bibinfo
  {author} {\bibfnamefont {A.}~\bibnamefont {Ernst}}, \ and\ \bibinfo {author}
  {\bibfnamefont {W.}~\bibnamefont {Wulfhekel}},\ }\bibfield  {title} {\enquote
  {\bibinfo {title} {Magnon dispersion in thin magnetic films},}\ }\href
  {\doibase 10.1088/0953-8984/26/39/394007} {\bibfield  {journal} {\bibinfo
  {journal} {Journal of Physics: Condensed Matter}\ }\textbf {\bibinfo {volume}
  {26}},\ \bibinfo {pages} {394007} (\bibinfo {year} {2014})}\BibitemShut
  {NoStop}%
\bibitem [{\citenamefont {Chuang}\ \emph {et~al.}(2014)\citenamefont {Chuang},
  \citenamefont {Zakeri}, \citenamefont {Ernst}, \citenamefont {Zhang},
  \citenamefont {Qin}, \citenamefont {Meng}, \citenamefont {Chen},\ and\
  \citenamefont {Kirschner}}]{Chuang.2014}%
  \BibitemOpen
  \bibfield  {author} {\bibinfo {author} {\bibfnamefont {T.-H.}\ \bibnamefont
  {Chuang}}, \bibinfo {author} {\bibfnamefont {K.}~\bibnamefont {Zakeri}},
  \bibinfo {author} {\bibfnamefont {A.}~\bibnamefont {Ernst}}, \bibinfo
  {author} {\bibfnamefont {Y.}~\bibnamefont {Zhang}}, \bibinfo {author}
  {\bibfnamefont {H.~J.}\ \bibnamefont {Qin}}, \bibinfo {author} {\bibfnamefont
  {Y.}~\bibnamefont {Meng}}, \bibinfo {author} {\bibfnamefont {Y.-J.}\
  \bibnamefont {Chen}}, \ and\ \bibinfo {author} {\bibfnamefont
  {J.}~\bibnamefont {Kirschner}},\ }\bibfield  {title} {\enquote {\bibinfo
  {title} {Magnetic properties and magnon excitations in {F}e(001) films grown
  on {I}r(001)},}\ }\href {\doibase 10.1103/PhysRevB.89.174404} {\bibfield
  {journal} {\bibinfo  {journal} {Physical Review B}\ }\textbf {\bibinfo
  {volume} {89}} (\bibinfo {year} {2014}),\
  10.1103/PhysRevB.89.174404}\BibitemShut {NoStop}%
\bibitem [{\citenamefont {Chumak}\ \emph {et~al.}(2015)\citenamefont {Chumak},
  \citenamefont {Vasyuchka}, \citenamefont {Serga},\ and\ \citenamefont
  {Hillebrands}}]{Chumak.2015}%
  \BibitemOpen
  \bibfield  {author} {\bibinfo {author} {\bibfnamefont {A.~V.}\ \bibnamefont
  {Chumak}}, \bibinfo {author} {\bibfnamefont {V.~I.}\ \bibnamefont
  {Vasyuchka}}, \bibinfo {author} {\bibfnamefont {A.~A.}\ \bibnamefont
  {Serga}}, \ and\ \bibinfo {author} {\bibfnamefont {B.}~\bibnamefont
  {Hillebrands}},\ }\bibfield  {title} {\enquote {\bibinfo {title} {Magnon
  spintronics},}\ }\href {\doibase 10.1038/nphys3347} {\bibfield  {journal}
  {\bibinfo  {journal} {Nature Physics}\ }\textbf {\bibinfo {volume} {11}},\
  \bibinfo {pages} {453--461} (\bibinfo {year} {2015})}\BibitemShut {NoStop}%
\bibitem [{\citenamefont {Hahn}\ \emph {et~al.}(2014)\citenamefont {Hahn},
  \citenamefont {Naletov}, \citenamefont {de~Loubens}, \citenamefont {Klein},
  \citenamefont {{d'Allivy Kelly}}, \citenamefont {Anane}, \citenamefont
  {Bernard}, \citenamefont {Jacquet}, \citenamefont {Bortolotti}, \citenamefont
  {Cros}, \citenamefont {Prieto},\ and\ \citenamefont {Mu{\~n}oz}}]{Hahn.2014}%
  \BibitemOpen
  \bibfield  {author} {\bibinfo {author} {\bibfnamefont {C.}~\bibnamefont
  {Hahn}}, \bibinfo {author} {\bibfnamefont {V.~V.}\ \bibnamefont {Naletov}},
  \bibinfo {author} {\bibfnamefont {G.}~\bibnamefont {de~Loubens}}, \bibinfo
  {author} {\bibfnamefont {O.}~\bibnamefont {Klein}}, \bibinfo {author}
  {\bibfnamefont {O.}~\bibnamefont {{d'Allivy Kelly}}}, \bibinfo {author}
  {\bibfnamefont {A.}~\bibnamefont {Anane}}, \bibinfo {author} {\bibfnamefont
  {R.}~\bibnamefont {Bernard}}, \bibinfo {author} {\bibfnamefont
  {E.}~\bibnamefont {Jacquet}}, \bibinfo {author} {\bibfnamefont
  {P.}~\bibnamefont {Bortolotti}}, \bibinfo {author} {\bibfnamefont
  {V.}~\bibnamefont {Cros}}, \bibinfo {author} {\bibfnamefont {J.~L.}\
  \bibnamefont {Prieto}}, \ and\ \bibinfo {author} {\bibfnamefont
  {M.}~\bibnamefont {Mu{\~n}oz}},\ }\bibfield  {title} {\enquote {\bibinfo
  {title} {Measurement of the intrinsic damping constant in individual
  nanodisks of $\mathrm{Y_3Fe_5O_{12}}$ and $\mathrm{Y_3Fe_5O_{12}|Pt}$},}\
  }\href {\doibase 10.1063/1.4871516} {\bibfield  {journal} {\bibinfo
  {journal} {Applied Physics Letters}\ }\textbf {\bibinfo {volume} {104}}
  (\bibinfo {year} {2014}),\ 10.1063/1.4871516}\BibitemShut {NoStop}%
\bibitem [{\citenamefont {Khitun}(2012)}]{Khitun.2012}%
  \BibitemOpen
  \bibfield  {author} {\bibinfo {author} {\bibfnamefont {A.}~\bibnamefont
  {Khitun}},\ }\bibfield  {title} {\enquote {\bibinfo {title} {Multi-frequency
  magnonic logic circuits for parallel data processing},}\ }\href {\doibase
  10.1063/1.3689011} {\bibfield  {journal} {\bibinfo  {journal} {Journal of
  Applied Physics}\ }\textbf {\bibinfo {volume} {111}} (\bibinfo {year}
  {2012}),\ 10.1063/1.3689011}\BibitemShut {NoStop}%
\bibitem [{\citenamefont {Khitun}, \citenamefont {Bao},\ and\ \citenamefont
  {Wang}(2010)}]{Khitun.2010}%
  \BibitemOpen
  \bibfield  {author} {\bibinfo {author} {\bibfnamefont {A.}~\bibnamefont
  {Khitun}}, \bibinfo {author} {\bibfnamefont {M.}~\bibnamefont {Bao}}, \ and\
  \bibinfo {author} {\bibfnamefont {K.~L.}\ \bibnamefont {Wang}},\ }\bibfield
  {title} {\enquote {\bibinfo {title} {Magnonic logic circuits},}\ }\href
  {\doibase 10.1088/0022-3727/43/26/264005} {\bibfield  {journal} {\bibinfo
  {journal} {Journal of Physics D: Applied Physics}\ }\textbf {\bibinfo
  {volume} {43}},\ \bibinfo {pages} {264005} (\bibinfo {year}
  {2010})}\BibitemShut {NoStop}%
\bibitem [{\citenamefont {Schneider}\ \emph
  {et~al.}(2008{\natexlab{a}})\citenamefont {Schneider}, \citenamefont {Serga},
  \citenamefont {Neumann}, \citenamefont {Hillebrands},\ and\ \citenamefont
  {Kostylev}}]{Schneider.2008b}%
  \BibitemOpen
  \bibfield  {author} {\bibinfo {author} {\bibfnamefont {T.}~\bibnamefont
  {Schneider}}, \bibinfo {author} {\bibfnamefont {A.~A.}\ \bibnamefont
  {Serga}}, \bibinfo {author} {\bibfnamefont {T.}~\bibnamefont {Neumann}},
  \bibinfo {author} {\bibfnamefont {B.}~\bibnamefont {Hillebrands}}, \ and\
  \bibinfo {author} {\bibfnamefont {M.~P.}\ \bibnamefont {Kostylev}},\
  }\bibfield  {title} {\enquote {\bibinfo {title} {Phase reciprocity of
  spin-wave excitation by a microstrip antenna},}\ }\href {\doibase
  10.1103/PhysRevB.77.214411} {\bibfield  {journal} {\bibinfo  {journal}
  {Physical Review B}\ }\textbf {\bibinfo {volume} {77}} (\bibinfo {year}
  {2008}{\natexlab{a}}),\ 10.1103/PhysRevB.77.214411}\BibitemShut {NoStop}%
\bibitem [{\citenamefont {Owens}, \citenamefont {Collins},\ and\ \citenamefont
  {Carter}(1985)}]{Owens.1985}%
  \BibitemOpen
  \bibfield  {author} {\bibinfo {author} {\bibfnamefont {J.~M.}\ \bibnamefont
  {Owens}}, \bibinfo {author} {\bibfnamefont {J.~H.}\ \bibnamefont {Collins}},
  \ and\ \bibinfo {author} {\bibfnamefont {R.~L.}\ \bibnamefont {Carter}},\
  }\bibfield  {title} {\enquote {\bibinfo {title} {System applications of
  magnetostatic wave devices},}\ }\href {\doibase 10.1007/BF01600088}
  {\bibfield  {journal} {\bibinfo  {journal} {Circuits, Systems, and Signal
  Processing}\ }\textbf {\bibinfo {volume} {4}},\ \bibinfo {pages} {317--334}
  (\bibinfo {year} {1985})}\BibitemShut {NoStop}%
\bibitem [{\citenamefont {Han}\ \emph {et~al.}(2023)\citenamefont {Han},
  \citenamefont {Cheng}, \citenamefont {Liu}, \citenamefont {Ohno},\ and\
  \citenamefont {Fukami}}]{Han.2023}%
  \BibitemOpen
  \bibfield  {author} {\bibinfo {author} {\bibfnamefont {J.}~\bibnamefont
  {Han}}, \bibinfo {author} {\bibfnamefont {R.}~\bibnamefont {Cheng}}, \bibinfo
  {author} {\bibfnamefont {L.}~\bibnamefont {Liu}}, \bibinfo {author}
  {\bibfnamefont {H.}~\bibnamefont {Ohno}}, \ and\ \bibinfo {author}
  {\bibfnamefont {S.}~\bibnamefont {Fukami}},\ }\bibfield  {title} {\enquote
  {\bibinfo {title} {Coherent antiferromagnetic spintronics},}\ }\href
  {\doibase 10.1038/s41563-023-01492-6} {\bibfield  {journal} {\bibinfo
  {journal} {Nature Materials}\ }\textbf {\bibinfo {volume} {22}},\ \bibinfo
  {pages} {684--695} (\bibinfo {year} {2023})}\BibitemShut {NoStop}%
\bibitem [{\citenamefont {Singh}\ \emph {et~al.}(2023)\citenamefont {Singh},
  \citenamefont {Kumar}, \citenamefont {Tyagi}, \citenamefont {Saxena},
  \citenamefont {Khan},\ and\ \citenamefont {Kumar}}]{Singh.2023}%
  \BibitemOpen
  \bibfield  {author} {\bibinfo {author} {\bibfnamefont {S.}~\bibnamefont
  {Singh}}, \bibinfo {author} {\bibfnamefont {V.}~\bibnamefont {Kumar}},
  \bibinfo {author} {\bibfnamefont {S.}~\bibnamefont {Tyagi}}, \bibinfo
  {author} {\bibfnamefont {N.}~\bibnamefont {Saxena}}, \bibinfo {author}
  {\bibfnamefont {Z.~H.}\ \bibnamefont {Khan}}, \ and\ \bibinfo {author}
  {\bibfnamefont {P.}~\bibnamefont {Kumar}},\ }\bibfield  {title} {\enquote
  {\bibinfo {title} {Room temperature ferromagnetism in metal oxides for
  spintronics: a comprehensive review},}\ }\href {\doibase
  10.1007/s11082-022-04325-z} {\bibfield  {journal} {\bibinfo  {journal}
  {Optical and Quantum Electronics}\ }\textbf {\bibinfo {volume} {55}},\
  \bibinfo {pages} {1--41} (\bibinfo {year} {2023})}\BibitemShut {NoStop}%
\bibitem [{\citenamefont {Zhang}\ \emph {et~al.}(2023)\citenamefont {Zhang},
  \citenamefont {Feng}, \citenamefont {Zheng}, \citenamefont {Zhang},
  \citenamefont {Lin}, \citenamefont {Sun}, \citenamefont {Wang}, \citenamefont
  {Wang}, \citenamefont {Wei}, \citenamefont {Vallobra}, \citenamefont {He},
  \citenamefont {Wang}, \citenamefont {Chen}, \citenamefont {Zhang},
  \citenamefont {Xu},\ and\ \citenamefont {Zhao}}]{Zhang.2023}%
  \BibitemOpen
  \bibfield  {author} {\bibinfo {author} {\bibfnamefont {Y.}~\bibnamefont
  {Zhang}}, \bibinfo {author} {\bibfnamefont {X.}~\bibnamefont {Feng}},
  \bibinfo {author} {\bibfnamefont {Z.}~\bibnamefont {Zheng}}, \bibinfo
  {author} {\bibfnamefont {Z.}~\bibnamefont {Zhang}}, \bibinfo {author}
  {\bibfnamefont {K.}~\bibnamefont {Lin}}, \bibinfo {author} {\bibfnamefont
  {X.}~\bibnamefont {Sun}}, \bibinfo {author} {\bibfnamefont {G.}~\bibnamefont
  {Wang}}, \bibinfo {author} {\bibfnamefont {J.}~\bibnamefont {Wang}}, \bibinfo
  {author} {\bibfnamefont {J.}~\bibnamefont {Wei}}, \bibinfo {author}
  {\bibfnamefont {P.}~\bibnamefont {Vallobra}}, \bibinfo {author}
  {\bibfnamefont {Y.}~\bibnamefont {He}}, \bibinfo {author} {\bibfnamefont
  {Z.}~\bibnamefont {Wang}}, \bibinfo {author} {\bibfnamefont {L.}~\bibnamefont
  {Chen}}, \bibinfo {author} {\bibfnamefont {K.}~\bibnamefont {Zhang}},
  \bibinfo {author} {\bibfnamefont {Y.}~\bibnamefont {Xu}}, \ and\ \bibinfo
  {author} {\bibfnamefont {W.}~\bibnamefont {Zhao}},\ }\bibfield  {title}
  {\enquote {\bibinfo {title} {Ferrimagnets for spintronic devices: From
  materials to applications},}\ }\href {\doibase 10.1063/5.0104618} {\bibfield
  {journal} {\bibinfo  {journal} {Applied Physics Reviews}\ }\textbf {\bibinfo
  {volume} {10}} (\bibinfo {year} {2023}),\ 10.1063/5.0104618}\BibitemShut
  {NoStop}%
\bibitem [{\citenamefont {Vilsmeier}, \citenamefont {Riedel},\ and\
  \citenamefont {Back}()}]{Vilsmeier.23.03.2024}%
  \BibitemOpen
  \bibfield  {author} {\bibinfo {author} {\bibfnamefont {F.}~\bibnamefont
  {Vilsmeier}}, \bibinfo {author} {\bibfnamefont {C.}~\bibnamefont {Riedel}}, \
  and\ \bibinfo {author} {\bibfnamefont {C.~H.}\ \bibnamefont {Back}},\ }\href
  {http://arxiv.org/pdf/2403.15840} {\enquote {\bibinfo {title} {Spatial
  control of hybridization-induced spin-wave transmission stop band},}\
  }\BibitemShut {NoStop}%
\bibitem [{\citenamefont {Perzlmaier}, \citenamefont {Woltersdorf},\ and\
  \citenamefont {Back}(2008)}]{Perzlmaier.2008}%
  \BibitemOpen
  \bibfield  {author} {\bibinfo {author} {\bibfnamefont {K.}~\bibnamefont
  {Perzlmaier}}, \bibinfo {author} {\bibfnamefont {G.}~\bibnamefont
  {Woltersdorf}}, \ and\ \bibinfo {author} {\bibfnamefont {C.~H.}\ \bibnamefont
  {Back}},\ }\bibfield  {title} {\enquote {\bibinfo {title} {Observation of the
  propagation and interference of spin waves in ferromagnetic thin films},}\
  }\href {\doibase 10.1103/physrevb.77.054425} {\bibfield  {journal} {\bibinfo
  {journal} {Physical Review B}\ }\textbf {\bibinfo {volume} {77}} (\bibinfo
  {year} {2008}),\ 10.1103/physrevb.77.054425}\BibitemShut {NoStop}%
\bibitem [{\citenamefont {Farle}, \citenamefont {Silva},\ and\ \citenamefont
  {Woltersdorf}(2013)}]{Farle.2013}%
  \BibitemOpen
  \bibfield  {author} {\bibinfo {author} {\bibfnamefont {M.}~\bibnamefont
  {Farle}}, \bibinfo {author} {\bibfnamefont {T.}~\bibnamefont {Silva}}, \ and\
  \bibinfo {author} {\bibfnamefont {G.}~\bibnamefont {Woltersdorf}},\
  }\bibfield  {title} {\enquote {\bibinfo {title} {Spin dynamics in the time
  and frequency domain},}\ }\href {\doibase
  10.1007/978-3-642-32042-2{\textunderscore }2} {\bibfield  {journal} {\bibinfo
   {journal} {Magnetic Nanostructures}\ }\textbf {\bibinfo {volume} {246}},\
  \bibinfo {pages} {37--83} (\bibinfo {year} {2013})}\BibitemShut {NoStop}%
\bibitem [{\citenamefont {Au}\ \emph {et~al.}(2011)\citenamefont {Au},
  \citenamefont {Davison}, \citenamefont {Ahmad}, \citenamefont {Keatley},
  \citenamefont {Hicken},\ and\ \citenamefont {Kruglyak}}]{Au.2011}%
  \BibitemOpen
  \bibfield  {author} {\bibinfo {author} {\bibfnamefont {Y.}~\bibnamefont
  {Au}}, \bibinfo {author} {\bibfnamefont {T.}~\bibnamefont {Davison}},
  \bibinfo {author} {\bibfnamefont {E.}~\bibnamefont {Ahmad}}, \bibinfo
  {author} {\bibfnamefont {P.~S.}\ \bibnamefont {Keatley}}, \bibinfo {author}
  {\bibfnamefont {R.~J.}\ \bibnamefont {Hicken}}, \ and\ \bibinfo {author}
  {\bibfnamefont {V.~V.}\ \bibnamefont {Kruglyak}},\ }\bibfield  {title}
  {\enquote {\bibinfo {title} {Excitation of propagating spin waves with global
  uniform microwave fields},}\ }\href {\doibase 10.1063/1.3571444} {\bibfield
  {journal} {\bibinfo  {journal} {Applied Physics Letters}\ }\textbf {\bibinfo
  {volume} {98}} (\bibinfo {year} {2011}),\ 10.1063/1.3571444}\BibitemShut
  {NoStop}%
\bibitem [{\citenamefont {Bauer}\ \emph {et~al.}(2014)\citenamefont {Bauer},
  \citenamefont {Chauleau}, \citenamefont {Woltersdorf},\ and\ \citenamefont
  {Back}}]{Bauer.2014}%
  \BibitemOpen
  \bibfield  {author} {\bibinfo {author} {\bibfnamefont {H.~G.}\ \bibnamefont
  {Bauer}}, \bibinfo {author} {\bibfnamefont {J.-Y.}\ \bibnamefont {Chauleau}},
  \bibinfo {author} {\bibfnamefont {G.}~\bibnamefont {Woltersdorf}}, \ and\
  \bibinfo {author} {\bibfnamefont {C.~H.}\ \bibnamefont {Back}},\ }\bibfield
  {title} {\enquote {\bibinfo {title} {Coupling of spinwave modes in wire
  structures},}\ }\href {\doibase 10.1063/1.4868250} {\bibfield  {journal}
  {\bibinfo  {journal} {Applied Physics Letters}\ }\textbf {\bibinfo {volume}
  {104}},\ \bibinfo {pages} {102404} (\bibinfo {year} {2014})}\BibitemShut
  {NoStop}%
\bibitem [{\citenamefont {Stigloher}\ \emph {et~al.}(2018)\citenamefont
  {Stigloher}, \citenamefont {Taniguchi}, \citenamefont {K{\"o}rner},
  \citenamefont {Decker}, \citenamefont {Moriyama}, \citenamefont {Ono},\ and\
  \citenamefont {Back}}]{Stigloher.2018}%
  \BibitemOpen
  \bibfield  {author} {\bibinfo {author} {\bibfnamefont {J.}~\bibnamefont
  {Stigloher}}, \bibinfo {author} {\bibfnamefont {T.}~\bibnamefont
  {Taniguchi}}, \bibinfo {author} {\bibfnamefont {H.~S.}\ \bibnamefont
  {K{\"o}rner}}, \bibinfo {author} {\bibfnamefont {M.}~\bibnamefont {Decker}},
  \bibinfo {author} {\bibfnamefont {T.}~\bibnamefont {Moriyama}}, \bibinfo
  {author} {\bibfnamefont {T.}~\bibnamefont {Ono}}, \ and\ \bibinfo {author}
  {\bibfnamefont {C.~H.}\ \bibnamefont {Back}},\ }\bibfield  {title} {\enquote
  {\bibinfo {title} {Observation of a goos-h{\"a}nchen-like phase shift for
  magnetostatic spin waves},}\ }\href {\doibase 10.1103/physrevlett.121.137201}
  {\bibfield  {journal} {\bibinfo  {journal} {Physical review letters}\
  }\textbf {\bibinfo {volume} {121}},\ \bibinfo {pages} {137201} (\bibinfo
  {year} {2018})}\BibitemShut {NoStop}%
\bibitem [{\citenamefont {Hillebrands}(1999)}]{Hillebrands.1999}%
  \BibitemOpen
  \bibfield  {author} {\bibinfo {author} {\bibfnamefont {B.}~\bibnamefont
  {Hillebrands}},\ }\bibfield  {title} {\enquote {\bibinfo {title} {Progress in
  multipass tandem fabry--perot interferometry: I. a fully automated, easy to
  use, self-aligning spectrometer with increased stability and flexibility},}\
  }\href {\doibase 10.1063/1.1149637} {\bibfield  {journal} {\bibinfo
  {journal} {Review of Scientific Instruments}\ }\textbf {\bibinfo {volume}
  {70}},\ \bibinfo {pages} {1589--1598} (\bibinfo {year} {1999})}\BibitemShut
  {NoStop}%
\bibitem [{\citenamefont {Demokritov}(2001)}]{Demokritov.2001}%
  \BibitemOpen
  \bibfield  {author} {\bibinfo {author} {\bibfnamefont {S.}~\bibnamefont
  {Demokritov}},\ }\bibfield  {title} {\enquote {\bibinfo {title} {Brillouin
  light scattering studies of confined spin waves: linear and nonlinear
  confinement},}\ }\href {\doibase 10.1016/s0370-1573(00)00116-2} {\bibfield
  {journal} {\bibinfo  {journal} {Physics Reports}\ }\textbf {\bibinfo {volume}
  {348}},\ \bibinfo {pages} {441--489} (\bibinfo {year} {2001})}\BibitemShut
  {NoStop}%
\bibitem [{\citenamefont {Warwick}\ \emph {et~al.}(1998)\citenamefont
  {Warwick}, \citenamefont {Franck}, \citenamefont {Kortright}, \citenamefont
  {Meigs}, \citenamefont {Moronne}, \citenamefont {Myneni}, \citenamefont
  {Rotenberg}, \citenamefont {Seal}, \citenamefont {Steele}, \citenamefont
  {Ade}, \citenamefont {Garcia}, \citenamefont {Cerasari}, \citenamefont
  {Denlinger}, \citenamefont {Hayakawa}, \citenamefont {Hitchcock},
  \citenamefont {Tyliszczak}, \citenamefont {Kikuma}, \citenamefont {Rightor},
  \citenamefont {Shin},\ and\ \citenamefont {Tonner}}]{Warwick.1998}%
  \BibitemOpen
  \bibfield  {author} {\bibinfo {author} {\bibfnamefont {T.}~\bibnamefont
  {Warwick}}, \bibinfo {author} {\bibfnamefont {K.}~\bibnamefont {Franck}},
  \bibinfo {author} {\bibfnamefont {J.~B.}\ \bibnamefont {Kortright}}, \bibinfo
  {author} {\bibfnamefont {G.}~\bibnamefont {Meigs}}, \bibinfo {author}
  {\bibfnamefont {M.}~\bibnamefont {Moronne}}, \bibinfo {author} {\bibfnamefont
  {S.}~\bibnamefont {Myneni}}, \bibinfo {author} {\bibfnamefont
  {E.}~\bibnamefont {Rotenberg}}, \bibinfo {author} {\bibfnamefont
  {S.}~\bibnamefont {Seal}}, \bibinfo {author} {\bibfnamefont {W.~F.}\
  \bibnamefont {Steele}}, \bibinfo {author} {\bibfnamefont {H.}~\bibnamefont
  {Ade}}, \bibinfo {author} {\bibfnamefont {A.}~\bibnamefont {Garcia}},
  \bibinfo {author} {\bibfnamefont {S.}~\bibnamefont {Cerasari}}, \bibinfo
  {author} {\bibfnamefont {J.}~\bibnamefont {Denlinger}}, \bibinfo {author}
  {\bibfnamefont {S.}~\bibnamefont {Hayakawa}}, \bibinfo {author}
  {\bibfnamefont {A.~P.}\ \bibnamefont {Hitchcock}}, \bibinfo {author}
  {\bibfnamefont {T.}~\bibnamefont {Tyliszczak}}, \bibinfo {author}
  {\bibfnamefont {J.}~\bibnamefont {Kikuma}}, \bibinfo {author} {\bibfnamefont
  {E.~G.}\ \bibnamefont {Rightor}}, \bibinfo {author} {\bibfnamefont {H.-J.}\
  \bibnamefont {Shin}}, \ and\ \bibinfo {author} {\bibfnamefont {B.~P.}\
  \bibnamefont {Tonner}},\ }\bibfield  {title} {\enquote {\bibinfo {title} {A
  scanning transmission x-ray microscope for materials science
  spectromicroscopy at the advanced light source},}\ }\href {\doibase
  10.1063/1.1149041} {\bibfield  {journal} {\bibinfo  {journal} {Review of
  Scientific Instruments}\ }\textbf {\bibinfo {volume} {69}},\ \bibinfo {pages}
  {2964--2973} (\bibinfo {year} {1998})}\BibitemShut {NoStop}%
\bibitem [{\citenamefont {Sluka}\ \emph {et~al.}(2019)\citenamefont {Sluka},
  \citenamefont {Schneider}, \citenamefont {Gallardo}, \citenamefont
  {K{\'a}kay}, \citenamefont {Weigand}, \citenamefont {Warnatz}, \citenamefont
  {Mattheis}, \citenamefont {Rold{\'a}n-Molina}, \citenamefont {Landeros},
  \citenamefont {Tiberkevich}, \citenamefont {Slavin}, \citenamefont
  {Sch{\"u}tz}, \citenamefont {Erbe}, \citenamefont {Deac}, \citenamefont
  {Lindner}, \citenamefont {Raabe}, \citenamefont {Fassbender},\ and\
  \citenamefont {Wintz}}]{Sluka.2019}%
  \BibitemOpen
  \bibfield  {author} {\bibinfo {author} {\bibfnamefont {V.}~\bibnamefont
  {Sluka}}, \bibinfo {author} {\bibfnamefont {T.}~\bibnamefont {Schneider}},
  \bibinfo {author} {\bibfnamefont {R.~A.}\ \bibnamefont {Gallardo}}, \bibinfo
  {author} {\bibfnamefont {A.}~\bibnamefont {K{\'a}kay}}, \bibinfo {author}
  {\bibfnamefont {M.}~\bibnamefont {Weigand}}, \bibinfo {author} {\bibfnamefont
  {T.}~\bibnamefont {Warnatz}}, \bibinfo {author} {\bibfnamefont
  {R.}~\bibnamefont {Mattheis}}, \bibinfo {author} {\bibfnamefont
  {A.}~\bibnamefont {Rold{\'a}n-Molina}}, \bibinfo {author} {\bibfnamefont
  {P.}~\bibnamefont {Landeros}}, \bibinfo {author} {\bibfnamefont
  {V.}~\bibnamefont {Tiberkevich}}, \bibinfo {author} {\bibfnamefont
  {A.}~\bibnamefont {Slavin}}, \bibinfo {author} {\bibfnamefont
  {G.}~\bibnamefont {Sch{\"u}tz}}, \bibinfo {author} {\bibfnamefont
  {A.}~\bibnamefont {Erbe}}, \bibinfo {author} {\bibfnamefont {A.}~\bibnamefont
  {Deac}}, \bibinfo {author} {\bibfnamefont {J.}~\bibnamefont {Lindner}},
  \bibinfo {author} {\bibfnamefont {J.}~\bibnamefont {Raabe}}, \bibinfo
  {author} {\bibfnamefont {J.}~\bibnamefont {Fassbender}}, \ and\ \bibinfo
  {author} {\bibfnamefont {S.}~\bibnamefont {Wintz}},\ }\bibfield  {title}
  {\enquote {\bibinfo {title} {Emission and propagation of 1d and 2d spin waves
  with nanoscale wavelengths in anisotropic spin textures},}\ }\href {\doibase
  10.1038/s41565-019-0383-4} {\bibfield  {journal} {\bibinfo  {journal} {Nature
  Nanotechnology}\ }\textbf {\bibinfo {volume} {14}},\ \bibinfo {pages}
  {328--333} (\bibinfo {year} {2019})}\BibitemShut {NoStop}%
\bibitem [{\citenamefont {Bertelli}\ \emph {et~al.}(2020)\citenamefont
  {Bertelli}, \citenamefont {Carmiggelt}, \citenamefont {Yu}, \citenamefont
  {Simon}, \citenamefont {Pothoven}, \citenamefont {Bauer}, \citenamefont
  {Blanter}, \citenamefont {Aarts},\ and\ \citenamefont {{van der
  Sar}}}]{Bertelli.2020}%
  \BibitemOpen
  \bibfield  {author} {\bibinfo {author} {\bibfnamefont {I.}~\bibnamefont
  {Bertelli}}, \bibinfo {author} {\bibfnamefont {J.~J.}\ \bibnamefont
  {Carmiggelt}}, \bibinfo {author} {\bibfnamefont {T.}~\bibnamefont {Yu}},
  \bibinfo {author} {\bibfnamefont {B.~G.}\ \bibnamefont {Simon}}, \bibinfo
  {author} {\bibfnamefont {C.~C.}\ \bibnamefont {Pothoven}}, \bibinfo {author}
  {\bibfnamefont {G.~E.~W.}\ \bibnamefont {Bauer}}, \bibinfo {author}
  {\bibfnamefont {Y.~M.}\ \bibnamefont {Blanter}}, \bibinfo {author}
  {\bibfnamefont {J.}~\bibnamefont {Aarts}}, \ and\ \bibinfo {author}
  {\bibfnamefont {T.}~\bibnamefont {{van der Sar}}},\ }\bibfield  {title}
  {\enquote {\bibinfo {title} {Magnetic resonance imaging of spin-wave
  transport and interference in a magnetic insulator},}\ }\href {\doibase
  10.1126/sciadv.abd3556} {\bibfield  {journal} {\bibinfo  {journal} {Science
  advances}\ }\textbf {\bibinfo {volume} {6}} (\bibinfo {year} {2020}),\
  10.1126/sciadv.abd3556}\BibitemShut {NoStop}%
\bibitem [{\citenamefont {Purser}\ \emph {et~al.}(2020)\citenamefont {Purser},
  \citenamefont {Bhallamudi}, \citenamefont {Guo}, \citenamefont {Page},
  \citenamefont {Guo}, \citenamefont {Fuchs},\ and\ \citenamefont
  {Hammel}}]{Purser.2020}%
  \BibitemOpen
  \bibfield  {author} {\bibinfo {author} {\bibfnamefont {C.~M.}\ \bibnamefont
  {Purser}}, \bibinfo {author} {\bibfnamefont {V.~P.}\ \bibnamefont
  {Bhallamudi}}, \bibinfo {author} {\bibfnamefont {F.}~\bibnamefont {Guo}},
  \bibinfo {author} {\bibfnamefont {M.~R.}\ \bibnamefont {Page}}, \bibinfo
  {author} {\bibfnamefont {Q.}~\bibnamefont {Guo}}, \bibinfo {author}
  {\bibfnamefont {G.~D.}\ \bibnamefont {Fuchs}}, \ and\ \bibinfo {author}
  {\bibfnamefont {P.~C.}\ \bibnamefont {Hammel}},\ }\bibfield  {title}
  {\enquote {\bibinfo {title} {Spinwave detection by nitrogen-vacancy centers
  in diamond as a function of probe--sample separation},}\ }\href {\doibase
  10.1063/1.5141921} {\bibfield  {journal} {\bibinfo  {journal} {Applied
  Physics Letters}\ }\textbf {\bibinfo {volume} {116}} (\bibinfo {year}
  {2020}),\ 10.1063/1.5141921}\BibitemShut {NoStop}%
\bibitem [{\citenamefont {Koerner}\ \emph {et~al.}(2022)\citenamefont
  {Koerner}, \citenamefont {Dreyer}, \citenamefont {Wagener}, \citenamefont
  {Liebing}, \citenamefont {Bauer},\ and\ \citenamefont
  {Woltersdorf}}]{Koerner.2022}%
  \BibitemOpen
  \bibfield  {author} {\bibinfo {author} {\bibfnamefont {C.}~\bibnamefont
  {Koerner}}, \bibinfo {author} {\bibfnamefont {R.}~\bibnamefont {Dreyer}},
  \bibinfo {author} {\bibfnamefont {M.}~\bibnamefont {Wagener}}, \bibinfo
  {author} {\bibfnamefont {N.}~\bibnamefont {Liebing}}, \bibinfo {author}
  {\bibfnamefont {H.~G.}\ \bibnamefont {Bauer}}, \ and\ \bibinfo {author}
  {\bibfnamefont {G.}~\bibnamefont {Woltersdorf}},\ }\bibfield  {title}
  {\enquote {\bibinfo {title} {Frequency multiplication by collective nanoscale
  spin-wave dynamics},}\ }\href {\doibase 10.1126/science.abm6044} {\bibfield
  {journal} {\bibinfo  {journal} {Science (New York, N.Y.)}\ }\textbf {\bibinfo
  {volume} {375}},\ \bibinfo {pages} {1165--1169} (\bibinfo {year}
  {2022})}\BibitemShut {NoStop}%
\bibitem [{\citenamefont {Tetienne}\ \emph {et~al.}(2013)\citenamefont
  {Tetienne}, \citenamefont {Hingant}, \citenamefont {Rondin}, \citenamefont
  {Cavaill{\`e}s}, \citenamefont {Mayer}, \citenamefont {Dantelle},
  \citenamefont {Gacoin}, \citenamefont {Wrachtrup}, \citenamefont {Roch},\
  and\ \citenamefont {Jacques}}]{Tetienne.2013}%
  \BibitemOpen
  \bibfield  {author} {\bibinfo {author} {\bibfnamefont {J.-P.}\ \bibnamefont
  {Tetienne}}, \bibinfo {author} {\bibfnamefont {T.}~\bibnamefont {Hingant}},
  \bibinfo {author} {\bibfnamefont {L.}~\bibnamefont {Rondin}}, \bibinfo
  {author} {\bibfnamefont {A.}~\bibnamefont {Cavaill{\`e}s}}, \bibinfo {author}
  {\bibfnamefont {L.}~\bibnamefont {Mayer}}, \bibinfo {author} {\bibfnamefont
  {G.}~\bibnamefont {Dantelle}}, \bibinfo {author} {\bibfnamefont
  {T.}~\bibnamefont {Gacoin}}, \bibinfo {author} {\bibfnamefont
  {J.}~\bibnamefont {Wrachtrup}}, \bibinfo {author} {\bibfnamefont {J.-F.}\
  \bibnamefont {Roch}}, \ and\ \bibinfo {author} {\bibfnamefont
  {V.}~\bibnamefont {Jacques}},\ }\bibfield  {title} {\enquote {\bibinfo
  {title} {Spin relaxometry of single nitrogen-vacancy defects in diamond
  nanocrystals for magnetic noise sensing},}\ }\href {\doibase
  10.1103/PhysRevB.87.235436} {\bibfield  {journal} {\bibinfo  {journal}
  {Physical Review B}\ }\textbf {\bibinfo {volume} {87}} (\bibinfo {year}
  {2013}),\ 10.1103/PhysRevB.87.235436}\BibitemShut {NoStop}%
\bibitem [{\citenamefont {Maletinsky}\ \emph {et~al.}(2012)\citenamefont
  {Maletinsky}, \citenamefont {Hong}, \citenamefont {Grinolds}, \citenamefont
  {Hausmann}, \citenamefont {Lukin}, \citenamefont {Walsworth}, \citenamefont
  {Loncar},\ and\ \citenamefont {Yacoby}}]{Maletinsky.2012}%
  \BibitemOpen
  \bibfield  {author} {\bibinfo {author} {\bibfnamefont {P.}~\bibnamefont
  {Maletinsky}}, \bibinfo {author} {\bibfnamefont {S.}~\bibnamefont {Hong}},
  \bibinfo {author} {\bibfnamefont {M.~S.}\ \bibnamefont {Grinolds}}, \bibinfo
  {author} {\bibfnamefont {B.}~\bibnamefont {Hausmann}}, \bibinfo {author}
  {\bibfnamefont {M.~D.}\ \bibnamefont {Lukin}}, \bibinfo {author}
  {\bibfnamefont {R.~L.}\ \bibnamefont {Walsworth}}, \bibinfo {author}
  {\bibfnamefont {M.}~\bibnamefont {Loncar}}, \ and\ \bibinfo {author}
  {\bibfnamefont {A.}~\bibnamefont {Yacoby}},\ }\bibfield  {title} {\enquote
  {\bibinfo {title} {A robust scanning diamond sensor for nanoscale imaging
  with single nitrogen-vacancy centres},}\ }\href {\doibase
  10.1038/nnano.2012.50} {\bibfield  {journal} {\bibinfo  {journal} {Nature
  Nanotechnology}\ }\textbf {\bibinfo {volume} {7}},\ \bibinfo {pages}
  {320--324} (\bibinfo {year} {2012})}\BibitemShut {NoStop}%
\bibitem [{\citenamefont {Appel}\ \emph {et~al.}(2015)\citenamefont {Appel},
  \citenamefont {Ganzhorn}, \citenamefont {Neu},\ and\ \citenamefont
  {Maletinsky}}]{Appel.2015}%
  \BibitemOpen
  \bibfield  {author} {\bibinfo {author} {\bibfnamefont {P.}~\bibnamefont
  {Appel}}, \bibinfo {author} {\bibfnamefont {M.}~\bibnamefont {Ganzhorn}},
  \bibinfo {author} {\bibfnamefont {E.}~\bibnamefont {Neu}}, \ and\ \bibinfo
  {author} {\bibfnamefont {P.}~\bibnamefont {Maletinsky}},\ }\bibfield  {title}
  {\enquote {\bibinfo {title} {Nanoscale microwave imaging with a single
  electron spin in diamond},}\ }\href {\doibase 10.1088/1367-2630/17/11/112001}
  {\bibfield  {journal} {\bibinfo  {journal} {New Journal of Physics}\ }\textbf
  {\bibinfo {volume} {17}},\ \bibinfo {pages} {112001} (\bibinfo {year}
  {2015})}\BibitemShut {NoStop}%
\bibitem [{\citenamefont {Rondin}\ \emph {et~al.}(2014)\citenamefont {Rondin},
  \citenamefont {Tetienne}, \citenamefont {Hingant}, \citenamefont {Roch},
  \citenamefont {Maletinsky},\ and\ \citenamefont {Jacques}}]{Rondin.2014}%
  \BibitemOpen
  \bibfield  {author} {\bibinfo {author} {\bibfnamefont {L.}~\bibnamefont
  {Rondin}}, \bibinfo {author} {\bibfnamefont {J.-P.}\ \bibnamefont
  {Tetienne}}, \bibinfo {author} {\bibfnamefont {T.}~\bibnamefont {Hingant}},
  \bibinfo {author} {\bibfnamefont {J.-F.}\ \bibnamefont {Roch}}, \bibinfo
  {author} {\bibfnamefont {P.}~\bibnamefont {Maletinsky}}, \ and\ \bibinfo
  {author} {\bibfnamefont {V.}~\bibnamefont {Jacques}},\ }\bibfield  {title}
  {\enquote {\bibinfo {title} {Magnetometry with nitrogen-vacancy defects in
  diamond},}\ }\href {\doibase 10.1088/0034-4885/77/5/056503} {\bibfield
  {journal} {\bibinfo  {journal} {Reports on progress in physics. Physical
  Society (Great Britain)}\ }\textbf {\bibinfo {volume} {77}},\ \bibinfo
  {pages} {056503} (\bibinfo {year} {2014})}\BibitemShut {NoStop}%
\bibitem [{\citenamefont {Balasubramanian}\ \emph {et~al.}(2008)\citenamefont
  {Balasubramanian}, \citenamefont {Chan}, \citenamefont {Kolesov},
  \citenamefont {Al-Hmoud}, \citenamefont {Tisler}, \citenamefont {Shin},
  \citenamefont {Kim}, \citenamefont {Wojcik}, \citenamefont {Hemmer},
  \citenamefont {Krueger}, \citenamefont {Hanke}, \citenamefont
  {Leitenstorfer}, \citenamefont {Bratschitsch}, \citenamefont {Jelezko},\ and\
  \citenamefont {Wrachtrup}}]{Balasubramanian.2008}%
  \BibitemOpen
  \bibfield  {author} {\bibinfo {author} {\bibfnamefont {G.}~\bibnamefont
  {Balasubramanian}}, \bibinfo {author} {\bibfnamefont {I.~Y.}\ \bibnamefont
  {Chan}}, \bibinfo {author} {\bibfnamefont {R.}~\bibnamefont {Kolesov}},
  \bibinfo {author} {\bibfnamefont {M.}~\bibnamefont {Al-Hmoud}}, \bibinfo
  {author} {\bibfnamefont {J.}~\bibnamefont {Tisler}}, \bibinfo {author}
  {\bibfnamefont {C.}~\bibnamefont {Shin}}, \bibinfo {author} {\bibfnamefont
  {C.}~\bibnamefont {Kim}}, \bibinfo {author} {\bibfnamefont {A.}~\bibnamefont
  {Wojcik}}, \bibinfo {author} {\bibfnamefont {P.~R.}\ \bibnamefont {Hemmer}},
  \bibinfo {author} {\bibfnamefont {A.}~\bibnamefont {Krueger}}, \bibinfo
  {author} {\bibfnamefont {T.}~\bibnamefont {Hanke}}, \bibinfo {author}
  {\bibfnamefont {A.}~\bibnamefont {Leitenstorfer}}, \bibinfo {author}
  {\bibfnamefont {R.}~\bibnamefont {Bratschitsch}}, \bibinfo {author}
  {\bibfnamefont {F.}~\bibnamefont {Jelezko}}, \ and\ \bibinfo {author}
  {\bibfnamefont {J.}~\bibnamefont {Wrachtrup}},\ }\bibfield  {title} {\enquote
  {\bibinfo {title} {Nanoscale imaging magnetometry with diamond spins under
  ambient conditions},}\ }\href {\doibase 10.1038/nature07278} {\bibfield
  {journal} {\bibinfo  {journal} {Nature}\ }\textbf {\bibinfo {volume} {455}},\
  \bibinfo {pages} {648--651} (\bibinfo {year} {2008})}\BibitemShut {NoStop}%
\bibitem [{\citenamefont {{van der Sar}}\ \emph {et~al.}(2015)\citenamefont
  {{van der Sar}}, \citenamefont {Casola}, \citenamefont {Walsworth},\ and\
  \citenamefont {Yacoby}}]{vanderSar.2015}%
  \BibitemOpen
  \bibfield  {author} {\bibinfo {author} {\bibfnamefont {T.}~\bibnamefont {{van
  der Sar}}}, \bibinfo {author} {\bibfnamefont {F.}~\bibnamefont {Casola}},
  \bibinfo {author} {\bibfnamefont {R.}~\bibnamefont {Walsworth}}, \ and\
  \bibinfo {author} {\bibfnamefont {A.}~\bibnamefont {Yacoby}},\ }\bibfield
  {title} {\enquote {\bibinfo {title} {Nanometre-scale probing of spin waves
  using single-electron spins},}\ }\href {\doibase 10.1038/ncomms8886}
  {\bibfield  {journal} {\bibinfo  {journal} {Nature communications}\ }\textbf
  {\bibinfo {volume} {6}},\ \bibinfo {pages} {7886} (\bibinfo {year}
  {2015})}\BibitemShut {NoStop}%
\bibitem [{\citenamefont {Bertelli}\ \emph {et~al.}(2021)\citenamefont
  {Bertelli}, \citenamefont {Simon}, \citenamefont {Yu}, \citenamefont {Aarts},
  \citenamefont {Bauer}, \citenamefont {Blanter},\ and\ \citenamefont {{van der
  Sar}}}]{Bertelli.2021}%
  \BibitemOpen
  \bibfield  {author} {\bibinfo {author} {\bibfnamefont {I.}~\bibnamefont
  {Bertelli}}, \bibinfo {author} {\bibfnamefont {B.~G.}\ \bibnamefont {Simon}},
  \bibinfo {author} {\bibfnamefont {T.}~\bibnamefont {Yu}}, \bibinfo {author}
  {\bibfnamefont {J.}~\bibnamefont {Aarts}}, \bibinfo {author} {\bibfnamefont
  {G.~E.~W.}\ \bibnamefont {Bauer}}, \bibinfo {author} {\bibfnamefont {Y.~M.}\
  \bibnamefont {Blanter}}, \ and\ \bibinfo {author} {\bibfnamefont
  {T.}~\bibnamefont {{van der Sar}}},\ }\bibfield  {title} {\enquote {\bibinfo
  {title} {Imaging spin--wave damping underneath metals using electron spins in
  diamond},}\ }\href {\doibase 10.1002/qute.202100094} {\bibfield  {journal}
  {\bibinfo  {journal} {Advanced Quantum Technologies}\ }\textbf {\bibinfo
  {volume} {4}} (\bibinfo {year} {2021}),\ 10.1002/qute.202100094}\BibitemShut
  {NoStop}%
\bibitem [{\citenamefont {Doherty}\ \emph {et~al.}(2013)\citenamefont
  {Doherty}, \citenamefont {Manson}, \citenamefont {Delaney}, \citenamefont
  {Jelezko}, \citenamefont {Wrachtrup},\ and\ \citenamefont
  {Hollenberg}}]{Doherty.2013}%
  \BibitemOpen
  \bibfield  {author} {\bibinfo {author} {\bibfnamefont {M.~W.}\ \bibnamefont
  {Doherty}}, \bibinfo {author} {\bibfnamefont {N.~B.}\ \bibnamefont {Manson}},
  \bibinfo {author} {\bibfnamefont {P.}~\bibnamefont {Delaney}}, \bibinfo
  {author} {\bibfnamefont {F.}~\bibnamefont {Jelezko}}, \bibinfo {author}
  {\bibfnamefont {J.}~\bibnamefont {Wrachtrup}}, \ and\ \bibinfo {author}
  {\bibfnamefont {L.~C.}\ \bibnamefont {Hollenberg}},\ }\bibfield  {title}
  {\enquote {\bibinfo {title} {The nitrogen-vacancy colour centre in
  diamond},}\ }\href {\doibase 10.1016/j.physrep.2013.02.001} {\bibfield
  {journal} {\bibinfo  {journal} {Physics Reports}\ }\textbf {\bibinfo {volume}
  {528}},\ \bibinfo {pages} {1--45} (\bibinfo {year} {2013})}\BibitemShut
  {NoStop}%
\bibitem [{\citenamefont {Andrich}\ \emph {et~al.}(2017)\citenamefont
  {Andrich}, \citenamefont {de~{las Casas}}, \citenamefont {Liu}, \citenamefont
  {Bretscher}, \citenamefont {Berman}, \citenamefont {Heremans}, \citenamefont
  {Nealey},\ and\ \citenamefont {Awschalom}}]{Andrich.2017}%
  \BibitemOpen
  \bibfield  {author} {\bibinfo {author} {\bibfnamefont {P.}~\bibnamefont
  {Andrich}}, \bibinfo {author} {\bibfnamefont {C.~F.}\ \bibnamefont {de~{las
  Casas}}}, \bibinfo {author} {\bibfnamefont {X.}~\bibnamefont {Liu}}, \bibinfo
  {author} {\bibfnamefont {H.~L.}\ \bibnamefont {Bretscher}}, \bibinfo {author}
  {\bibfnamefont {J.~R.}\ \bibnamefont {Berman}}, \bibinfo {author}
  {\bibfnamefont {F.~J.}\ \bibnamefont {Heremans}}, \bibinfo {author}
  {\bibfnamefont {P.~F.}\ \bibnamefont {Nealey}}, \ and\ \bibinfo {author}
  {\bibfnamefont {D.~D.}\ \bibnamefont {Awschalom}},\ }\bibfield  {title}
  {\enquote {\bibinfo {title} {Long-range spin wave mediated control of defect
  qubits in nanodiamonds},}\ }\href {\doibase 10.1038/s41534-017-0029-z}
  {\bibfield  {journal} {\bibinfo  {journal} {npj Quantum Information}\
  }\textbf {\bibinfo {volume} {3}} (\bibinfo {year} {2017}),\
  10.1038/s41534-017-0029-z}\BibitemShut {NoStop}%
\bibitem [{\citenamefont {Damon}\ and\ \citenamefont
  {Eshbach}(1961)}]{Damon.1961}%
  \BibitemOpen
  \bibfield  {author} {\bibinfo {author} {\bibfnamefont {R.~W.}\ \bibnamefont
  {Damon}}\ and\ \bibinfo {author} {\bibfnamefont {J.~R.}\ \bibnamefont
  {Eshbach}},\ }\bibfield  {title} {\enquote {\bibinfo {title} {Magnetostatic
  modes of a ferromagnet slab},}\ }\href {\doibase
  10.1016/0022-3697(61)90041-5} {\bibfield  {journal} {\bibinfo  {journal}
  {Journal of Physics and Chemistry of Solids}\ }\textbf {\bibinfo {volume}
  {19}},\ \bibinfo {pages} {308--320} (\bibinfo {year} {1961})}\BibitemShut
  {NoStop}%
\bibitem [{\citenamefont {Dreyer}\ \emph {et~al.}(2021)\citenamefont {Dreyer},
  \citenamefont {Liebing}, \citenamefont {Edwards}, \citenamefont
  {M{\"u}ller},\ and\ \citenamefont {Woltersdorf}}]{Dreyer.2021}%
  \BibitemOpen
  \bibfield  {author} {\bibinfo {author} {\bibfnamefont {R.}~\bibnamefont
  {Dreyer}}, \bibinfo {author} {\bibfnamefont {N.}~\bibnamefont {Liebing}},
  \bibinfo {author} {\bibfnamefont {E.~R.~J.}\ \bibnamefont {Edwards}},
  \bibinfo {author} {\bibfnamefont {A.}~\bibnamefont {M{\"u}ller}}, \ and\
  \bibinfo {author} {\bibfnamefont {G.}~\bibnamefont {Woltersdorf}},\
  }\bibfield  {title} {\enquote {\bibinfo {title} {Spin-wave localization and
  guiding by magnon band structure engineering in yttrium iron garnet},}\
  }\href {\doibase 10.1103/PhysRevMaterials.5.064411} {\bibfield  {journal}
  {\bibinfo  {journal} {Physical Review Materials}\ }\textbf {\bibinfo {volume}
  {5}} (\bibinfo {year} {2021}),\
  10.1103/PhysRevMaterials.5.064411}\BibitemShut {NoStop}%
\bibitem [{\citenamefont {Borst}\ \emph {et~al.}(2023)\citenamefont {Borst},
  \citenamefont {Vree}, \citenamefont {Lowther}, \citenamefont {Teepe},
  \citenamefont {Kurdi}, \citenamefont {Bertelli}, \citenamefont {Simon},
  \citenamefont {Blanter},\ and\ \citenamefont {{van der Sar}}}]{Borst.2023}%
  \BibitemOpen
  \bibfield  {author} {\bibinfo {author} {\bibfnamefont {M.}~\bibnamefont
  {Borst}}, \bibinfo {author} {\bibfnamefont {P.~H.}\ \bibnamefont {Vree}},
  \bibinfo {author} {\bibfnamefont {A.}~\bibnamefont {Lowther}}, \bibinfo
  {author} {\bibfnamefont {A.}~\bibnamefont {Teepe}}, \bibinfo {author}
  {\bibfnamefont {S.}~\bibnamefont {Kurdi}}, \bibinfo {author} {\bibfnamefont
  {I.}~\bibnamefont {Bertelli}}, \bibinfo {author} {\bibfnamefont {B.~G.}\
  \bibnamefont {Simon}}, \bibinfo {author} {\bibfnamefont {Y.~M.}\ \bibnamefont
  {Blanter}}, \ and\ \bibinfo {author} {\bibfnamefont {T.}~\bibnamefont {{van
  der Sar}}},\ }\bibfield  {title} {\enquote {\bibinfo {title} {Observation and
  control of hybrid spin-wave-meissner-current transport modes},}\ }\href
  {\doibase 10.1126/science.adj7576} {\bibfield  {journal} {\bibinfo  {journal}
  {Science (New York, N.Y.)}\ }\textbf {\bibinfo {volume} {382}},\ \bibinfo
  {pages} {430--434} (\bibinfo {year} {2023})}\BibitemShut {NoStop}%
\bibitem [{\citenamefont {Klingler}\ \emph {et~al.}(2017)\citenamefont
  {Klingler}, \citenamefont {Maier-Flaig}, \citenamefont {Dubs}, \citenamefont
  {Surzhenko}, \citenamefont {Gross}, \citenamefont {Huebl}, \citenamefont
  {Goennenwein},\ and\ \citenamefont {Weiler}}]{Klingler.2017}%
  \BibitemOpen
  \bibfield  {author} {\bibinfo {author} {\bibfnamefont {S.}~\bibnamefont
  {Klingler}}, \bibinfo {author} {\bibfnamefont {H.}~\bibnamefont
  {Maier-Flaig}}, \bibinfo {author} {\bibfnamefont {C.}~\bibnamefont {Dubs}},
  \bibinfo {author} {\bibfnamefont {O.}~\bibnamefont {Surzhenko}}, \bibinfo
  {author} {\bibfnamefont {R.}~\bibnamefont {Gross}}, \bibinfo {author}
  {\bibfnamefont {H.}~\bibnamefont {Huebl}}, \bibinfo {author} {\bibfnamefont
  {S.~T.~B.}\ \bibnamefont {Goennenwein}}, \ and\ \bibinfo {author}
  {\bibfnamefont {M.}~\bibnamefont {Weiler}},\ }\bibfield  {title} {\enquote
  {\bibinfo {title} {Gilbert damping of magnetostatic modes in a yttrium iron
  garnet sphere},}\ }\href {\doibase 10.1063/1.4977423} {\bibfield  {journal}
  {\bibinfo  {journal} {Applied Physics Letters}\ }\textbf {\bibinfo {volume}
  {110}} (\bibinfo {year} {2017}),\ 10.1063/1.4977423}\BibitemShut {NoStop}%
\bibitem [{\citenamefont {Osterkamp}\ \emph {et~al.}(2020)\citenamefont
  {Osterkamp}, \citenamefont {Balasubramanian}, \citenamefont {Wolff},
  \citenamefont {Teraji}, \citenamefont {Nesladek},\ and\ \citenamefont
  {Jelezko}}]{Osterkamp.2020}%
  \BibitemOpen
  \bibfield  {author} {\bibinfo {author} {\bibfnamefont {C.}~\bibnamefont
  {Osterkamp}}, \bibinfo {author} {\bibfnamefont {P.}~\bibnamefont
  {Balasubramanian}}, \bibinfo {author} {\bibfnamefont {G.}~\bibnamefont
  {Wolff}}, \bibinfo {author} {\bibfnamefont {T.}~\bibnamefont {Teraji}},
  \bibinfo {author} {\bibfnamefont {M.}~\bibnamefont {Nesladek}}, \ and\
  \bibinfo {author} {\bibfnamefont {F.}~\bibnamefont {Jelezko}},\ }\bibfield
  {title} {\enquote {\bibinfo {title} {Benchmark for synthesized diamond
  sensors based on isotopically engineered nitrogen--vacancy spin ensembles for
  magnetometry applications},}\ }\href {\doibase 10.1002/qute.202000074}
  {\bibfield  {journal} {\bibinfo  {journal} {Advanced Quantum Technologies}\
  }\textbf {\bibinfo {volume} {3}} (\bibinfo {year} {2020}),\
  10.1002/qute.202000074}\BibitemShut {NoStop}%
\bibitem [{\citenamefont {Schneider}\ \emph
  {et~al.}(2008{\natexlab{b}})\citenamefont {Schneider}, \citenamefont {Serga},
  \citenamefont {Leven}, \citenamefont {Hillebrands}, \citenamefont {Stamps},\
  and\ \citenamefont {Kostylev}}]{Schneider.2008}%
  \BibitemOpen
  \bibfield  {author} {\bibinfo {author} {\bibfnamefont {T.}~\bibnamefont
  {Schneider}}, \bibinfo {author} {\bibfnamefont {A.~A.}\ \bibnamefont
  {Serga}}, \bibinfo {author} {\bibfnamefont {B.}~\bibnamefont {Leven}},
  \bibinfo {author} {\bibfnamefont {B.}~\bibnamefont {Hillebrands}}, \bibinfo
  {author} {\bibfnamefont {R.~L.}\ \bibnamefont {Stamps}}, \ and\ \bibinfo
  {author} {\bibfnamefont {M.~P.}\ \bibnamefont {Kostylev}},\ }\bibfield
  {title} {\enquote {\bibinfo {title} {Realization of spin-wave logic gates},}\
  }\href {\doibase 10.1063/1.2834714} {\bibfield  {journal} {\bibinfo
  {journal} {Applied Physics Letters}\ }\textbf {\bibinfo {volume} {92}}
  (\bibinfo {year} {2008}{\natexlab{b}}),\ 10.1063/1.2834714}\BibitemShut
  {NoStop}%
\bibitem [{\citenamefont {Stancil}\ and\ \citenamefont
  {Prabhakar}(2009)}]{Stancil.2009}%
  \BibitemOpen
  \bibfield  {author} {\bibinfo {author} {\bibfnamefont {D.~D.}\ \bibnamefont
  {Stancil}}\ and\ \bibinfo {author} {\bibfnamefont {A.}~\bibnamefont
  {Prabhakar}},\ }\href@noop {} {\enquote {\bibinfo {title} {Spin waves: Theory
  and applications},}\ } (\bibinfo {year} {2009})\BibitemShut {NoStop}%
\bibitem [{\citenamefont {Katsumata}\ \emph {et~al.}(2010)\citenamefont
  {Katsumata}, \citenamefont {Katori}, \citenamefont {Kimura}, \citenamefont
  {Narumi}, \citenamefont {Hagiwara},\ and\ \citenamefont
  {Kindo}}]{Katsumata.2010}%
  \BibitemOpen
  \bibfield  {author} {\bibinfo {author} {\bibfnamefont {K.}~\bibnamefont
  {Katsumata}}, \bibinfo {author} {\bibfnamefont {H.~A.}\ \bibnamefont
  {Katori}}, \bibinfo {author} {\bibfnamefont {S.}~\bibnamefont {Kimura}},
  \bibinfo {author} {\bibfnamefont {Y.}~\bibnamefont {Narumi}}, \bibinfo
  {author} {\bibfnamefont {M.}~\bibnamefont {Hagiwara}}, \ and\ \bibinfo
  {author} {\bibfnamefont {K.}~\bibnamefont {Kindo}},\ }\bibfield  {title}
  {\enquote {\bibinfo {title} {Phase transition of a triangular lattice ising
  antiferromagnet fei2},}\ }\href {\doibase 10.1103/PhysRevB.82.104402}
  {\bibfield  {journal} {\bibinfo  {journal} {Physical Review B}\ }\textbf
  {\bibinfo {volume} {82}} (\bibinfo {year} {2010}),\
  10.1103/PhysRevB.82.104402}\BibitemShut {NoStop}%
\bibitem [{\citenamefont {Joshi}\ \emph {et~al.}(2019)\citenamefont {Joshi},
  \citenamefont {Nusran}, \citenamefont {Tanatar}, \citenamefont {Cho},
  \citenamefont {Meier}, \citenamefont {Bud'ko}, \citenamefont {Canfield},\
  and\ \citenamefont {Prozorov}}]{Joshi.2019}%
  \BibitemOpen
  \bibfield  {author} {\bibinfo {author} {\bibfnamefont {K.~R.}\ \bibnamefont
  {Joshi}}, \bibinfo {author} {\bibfnamefont {N.~M.}\ \bibnamefont {Nusran}},
  \bibinfo {author} {\bibfnamefont {M.~A.}\ \bibnamefont {Tanatar}}, \bibinfo
  {author} {\bibfnamefont {K.}~\bibnamefont {Cho}}, \bibinfo {author}
  {\bibfnamefont {W.~R.}\ \bibnamefont {Meier}}, \bibinfo {author}
  {\bibfnamefont {S.~L.}\ \bibnamefont {Bud'ko}}, \bibinfo {author}
  {\bibfnamefont {P.~C.}\ \bibnamefont {Canfield}}, \ and\ \bibinfo {author}
  {\bibfnamefont {R.}~\bibnamefont {Prozorov}},\ }\bibfield  {title} {\enquote
  {\bibinfo {title} {Measuring the lower critical field of superconductors
  using nitrogen-vacancy centers in diamond optical magnetometry},}\ }\href
  {\doibase 10.1103/PhysRevApplied.11.014035} {\bibfield  {journal} {\bibinfo
  {journal} {Physical Review Applied}\ }\textbf {\bibinfo {volume} {11}}
  (\bibinfo {year} {2019}),\ 10.1103/PhysRevApplied.11.014035}\BibitemShut
  {NoStop}%
\bibitem [{\citenamefont {Ariyaratne}\ \emph {et~al.}(2018)\citenamefont
  {Ariyaratne}, \citenamefont {Bluvstein}, \citenamefont {Myers},\ and\
  \citenamefont {Jayich}}]{Ariyaratne.2018}%
  \BibitemOpen
  \bibfield  {author} {\bibinfo {author} {\bibfnamefont {A.}~\bibnamefont
  {Ariyaratne}}, \bibinfo {author} {\bibfnamefont {D.}~\bibnamefont
  {Bluvstein}}, \bibinfo {author} {\bibfnamefont {B.~A.}\ \bibnamefont
  {Myers}}, \ and\ \bibinfo {author} {\bibfnamefont {A.~C.~B.}\ \bibnamefont
  {Jayich}},\ }\bibfield  {title} {\enquote {\bibinfo {title} {Nanoscale
  electrical conductivity imaging using a nitrogen-vacancy center in
  diamond},}\ }\href {\doibase 10.1038/s41467-018-04798-1} {\bibfield
  {journal} {\bibinfo  {journal} {Nature communications}\ }\textbf {\bibinfo
  {volume} {9}},\ \bibinfo {pages} {2406} (\bibinfo {year} {2018})}\BibitemShut
  {NoStop}%
\bibitem [{\citenamefont {Finco}\ \emph {et~al.}(2021)\citenamefont {Finco},
  \citenamefont {Haykal}, \citenamefont {Tanos}, \citenamefont {Fabre},
  \citenamefont {Chouaieb}, \citenamefont {Akhtar}, \citenamefont
  {Robert-Philip}, \citenamefont {Legrand}, \citenamefont {Ajejas},
  \citenamefont {Bouzehouane}, \citenamefont {Reyren}, \citenamefont
  {Devolder}, \citenamefont {Adam}, \citenamefont {Kim}, \citenamefont {Cros},\
  and\ \citenamefont {Jacques}}]{Finco.2021}%
  \BibitemOpen
  \bibfield  {author} {\bibinfo {author} {\bibfnamefont {A.}~\bibnamefont
  {Finco}}, \bibinfo {author} {\bibfnamefont {A.}~\bibnamefont {Haykal}},
  \bibinfo {author} {\bibfnamefont {R.}~\bibnamefont {Tanos}}, \bibinfo
  {author} {\bibfnamefont {F.}~\bibnamefont {Fabre}}, \bibinfo {author}
  {\bibfnamefont {S.}~\bibnamefont {Chouaieb}}, \bibinfo {author}
  {\bibfnamefont {W.}~\bibnamefont {Akhtar}}, \bibinfo {author} {\bibfnamefont
  {I.}~\bibnamefont {Robert-Philip}}, \bibinfo {author} {\bibfnamefont
  {W.}~\bibnamefont {Legrand}}, \bibinfo {author} {\bibfnamefont
  {F.}~\bibnamefont {Ajejas}}, \bibinfo {author} {\bibfnamefont
  {K.}~\bibnamefont {Bouzehouane}}, \bibinfo {author} {\bibfnamefont
  {N.}~\bibnamefont {Reyren}}, \bibinfo {author} {\bibfnamefont
  {T.}~\bibnamefont {Devolder}}, \bibinfo {author} {\bibfnamefont {J.-P.}\
  \bibnamefont {Adam}}, \bibinfo {author} {\bibfnamefont {J.-V.}\ \bibnamefont
  {Kim}}, \bibinfo {author} {\bibfnamefont {V.}~\bibnamefont {Cros}}, \ and\
  \bibinfo {author} {\bibfnamefont {V.}~\bibnamefont {Jacques}},\ }\bibfield
  {title} {\enquote {\bibinfo {title} {Imaging non-collinear antiferromagnetic
  textures via single spin relaxometry},}\ }\href {\doibase
  10.1038/s41467-021-20995-x} {\bibfield  {journal} {\bibinfo  {journal}
  {Nature communications}\ }\textbf {\bibinfo {volume} {12}},\ \bibinfo {pages}
  {767} (\bibinfo {year} {2021})}\BibitemShut {NoStop}%
\bibitem [{\citenamefont {Gurevich}\ and\ \citenamefont
  {Melkov}(1996{\natexlab{b}})}]{Gurevich.1996b}%
  \BibitemOpen
  \bibfield  {author} {\bibinfo {author} {\bibfnamefont {A.~G.}\ \bibnamefont
  {Gurevich}}\ and\ \bibinfo {author} {\bibfnamefont {G.~A.}\ \bibnamefont
  {Melkov}},\ }\href@noop {} {\enquote {\bibinfo {title} {Magnetization
  oscillations and waves},}\ } (\bibinfo {year}
  {1996}{\natexlab{b}})\BibitemShut {NoStop}%
\end{thebibliography}%

\appendix

\section{Diamond substrate and YIG sample}
\label{A:sample}
\subsection{Diamond substrate}
We use a type IIa single-crystal diamond of size 0.5~mm $\times$ 0.5~mm $\times$ 100 $\mu$m grown by chemical vapour deposition, with Nitrogen content $<$ 1~ppm, surface roughness $<$ 2~nm Ra, and (100)-orientation (Applied Diamond). 
It was implanted with $\mathrm{^{15}N^+}$ ions at an energy of 4 keV, a dose of  $1.2 \times 10^{13}$ ions$/\mathrm{cm^2}$, and a tilt angle of 7$^\circ$ (performed by CuttingEdge Ions), resulting in an implantation depth of 9~nm.
Following this, it underwent annealing in a vacuum environment of $\sim 10^{-6}$~mbar. The annealing process involved a 45-minute ramp to 900$^\circ$C, followed by 3 hours at 900$^\circ$C~$\pm 10^\circ$C and a 2 h ramp to room temperature. 
To remove the graphitic layer that formed during the annealing, it was acid cleaned during 4 h by a boiling 3-acid mix consisting of 3 ml sulfuric acid ($\mathrm{H_2SO_4}$), 3 ml nitric acid ($\mathrm{HNO_3}$), and 3 ml perchloric acid  ($\mathrm{HClO_4}$).

\subsection{YIG sample}
We use a 200~nm thick YIG thin film grown on a gadolinium gallium garnet substrate via liquid phase epitaxy. The thin film posses a saturation magnetization $\mu_0 M_\mathrm{S} = 0.185$~mT and an exchange stiffness $A_\mathrm{ex} = 3.7 \cdot 10^{-12} \frac{\mathrm{J}}{\mathrm{m}}$. To excite spin waves a stripline S1 with a with of 5~$\mu$m was fabricated on top of the YIG film by optical lithography and electron beam evaporation of Ti(5~nm)/Au(100~nm). A second stripline S2 of Ti(5~nm)/Au(200~nm) with width 30~$\mu$m as well as a 150~$\mu$m thick $\mathrm{SiO_2}$ layer were fabricated on top of S1 by optical lithography and electron beam evaporation.

\section{Measurement setup}
\label{A:setup}

\subsection{NV center setup}
The setup used for the NV center measurements was an in-house built confocal microscope. The NV centers were optically excited by a 515~nm laser (iBeam-smart-515s, Toptica), which was focused to a diffraction-limited spot by an objective with a numerical aperture of 0.9 and a magnification of 63. Before the objective, a polarizer in combination with a $\lambda/2$-plate were used to rotate the polarization of the laser light, maximizing the contrast in the $\omega_\pm$.
The NV PL was gathered by the same objective, separated from the excitation light by a dichroic short pass mirror (cut-off wavelength 600~nm, Edmund Optics) and a long-pass filter (cut-on wavelength 590~nm). 
Microwaves for driving the NVs and spin waves were generated using a microwave generator (N5183A MXG, Agilent). To simultaneously send a microwave current through the pair of striplines S1 and S2 depicted in Fig.~\ref{fig:sample}~a), the microwave excitation was divided using a power combiner (ZFRSC-123-S+, Mini-Circuits). To measure the PL signal with a lock-in amplifier the microwave signal was modulated by a coaxial switch (139-ZASWA-2-50DRA, Mouser Electronics).
The PL photons were collected by an avalanche photodiode and, after passing a voltage amplifier, measured by the lock-in amplifier. Each data point was measured during 1~s at the 3th filter order. All measurements were conducted at room temperature.

\subsection{TR-MOKE setup}
The TR-MOKE measurements utilized a custom-built confocal microscope. A pulsed 800~nm Ti:Sa-Laser was employed. The beam then passed through a pellicle beam splitter with 92\% transmission. Given that the sample's magnetization would influence the light's polarization, a polarizing prism was employed for polarization component analysis. A Wollaston prism split the returning beam into two linearly polarized components, each directed towards a differential detector with two photodiodes. To synchronize the laser with the excitation, the laser's cavity length was adjusted to match the repetition rate with a global frequency reference. This adjustment was facilitated by a control loop incorporating a photodiode detecting generated pulses, mixed with the reference signal to produce a down-converted signal for use by a phase lock loop and proportional-integral-derivative controller. The reference signal, derived from a 10 MHz rubidium atomic clock underwent frequency multiplication to achieve the desired 80~MHz value. This reference signal was also utilized for an arbitrary waveform generator producing MW signals sent to the sample's stripline S1 (Fig.~\ref{fig:sample}~a)). The output of the lock-in is referred to as the Kerr signal $S_\mathrm{Kerr}$, while the summed signal of the two photodiodes is referred to as the topographic signal $S_\mathrm{Topo}$.

\section{Simulation method}
\label{A:simulation}
To simulate the PL signal resulting from the standing wave pattern of the total magnetic field above the YIG thin film, the time-averaged magnitude of total magnetic stray field perpendicular to the NV axis $<|B_{\mathrm{tot}_\perp}|>_t$ was calculated, where $B_\mathrm{tot_\perp}(t)$ is given by
\begin{equation}
    B_\mathrm{tot_\perp}(t)  = (B_\mathrm{sw}(t) + B_\mathrm{oe}(t))_\perp.
\end{equation}
The time dependent spin wave stray field $B_\mathrm{sw}(t)$ is given by \cite{Bertelli.2020}
\begin{equation}
    B_\mathrm{sw}(t) = - B_\mathrm{sw}\mathrm{Re}[e^{i(k_yy-\omega t)}(\hat{\mathbf{y}}+i\mathrm{sgn}(k_y)\hat{\mathbf{z}})],
\end{equation}
$B_\mathrm{sw} = \mu_0 m_0 (1+\mathrm{sgn}(k_y)\eta) |\mathbf{k}| d e^{-k_y z_0}$, where $\mathbf{k}$ is the wave vector, $\omega = 2\pi f$ is the frequency, $\eta$ is the degree of ellipticity, and $t$ is the time. For the parameters, we used $m_0 = 0.8 M_\mathrm{S}$ for the magnetization, $z_0 = 1~\mu$m for the distance between the NV layer and the YIG surface, and $\eta = 1$ for the ellipticity.
Furthermore, the time-dependent Oersted field of S2 is given by \cite{Appel.2015}
\begin{equation}
    \begin{aligned}
       B_\mathrm{oe}(t) &= -C\bigg (\mathrm{atan}(\frac{w-2x}{2z_0}) + \mathrm{atan}(\frac{w+2x}{2z_0}))\hat{\mathbf{x}} \\
       &+ \frac{1}{2} \big (-\mathrm{log}((w-2x)^2 + 4z_0^2) \\
       &+ \mathrm{log}((w+2x)^2 + 4z_0^2) \big ) \hat{\mathbf{z}} \bigg ) \mathrm{cos}(\omega t),
    \end{aligned}
\end{equation}
where $w = 30~\mu$m is the width of S2 and $C =  -(\mu_0 J h)/(2\pi w)$. Here, $J =  0.08~\mathrm{mA}/ (\mu\mathrm{m})^2$ is the microwave current density sent through S2, and $h = 250$~nm is the distance between the thin film surface and S2.

\section{Calulation of the dispersion}
\label{A:Dispersion}
The spin wave dispersion for free magnetization boundary conditions is given by \cite{Bertelli.2020}
\begin{equation}
    \begin{aligned}
        \omega &= \gamma \mu_0 M_\mathrm{S}\sqrt{\Omega_\mathrm{H} + \alpha_\mathrm{ex}k^2 + 1 - f(|k_y|)}\\
        &\cdot \sqrt{\Omega_\mathrm{H} + \alpha_\mathrm{ex}k^2 + (k_y^2/k^2)f(|k_y|)},
    \end{aligned}
\end{equation}
with $\Omega_\mathrm{H} = B \mathrm{cos}(\phi)/\mu_0 M_\mathrm{S}$, the exchange stiffness $\alpha_\mathrm{ex}$, the saturation magnetization $M_\mathrm{S}$, the angle $\phi = 35^\circ$, and
\begin{equation}
    f(|k_y|) = 1 - \frac{1}{|k_y|d} + \frac{1}{|k_y|d}\mathrm{exp}(-|k_y|d),
\end{equation}
where $d = 200$~nm is the YIG thickness.

\section{Wavefront measurements in dependence of the external field}
\label{A:2D wavefronts}
\begin{figure}[htb!]
    \centering
    \includegraphics[width=0.45\textwidth]{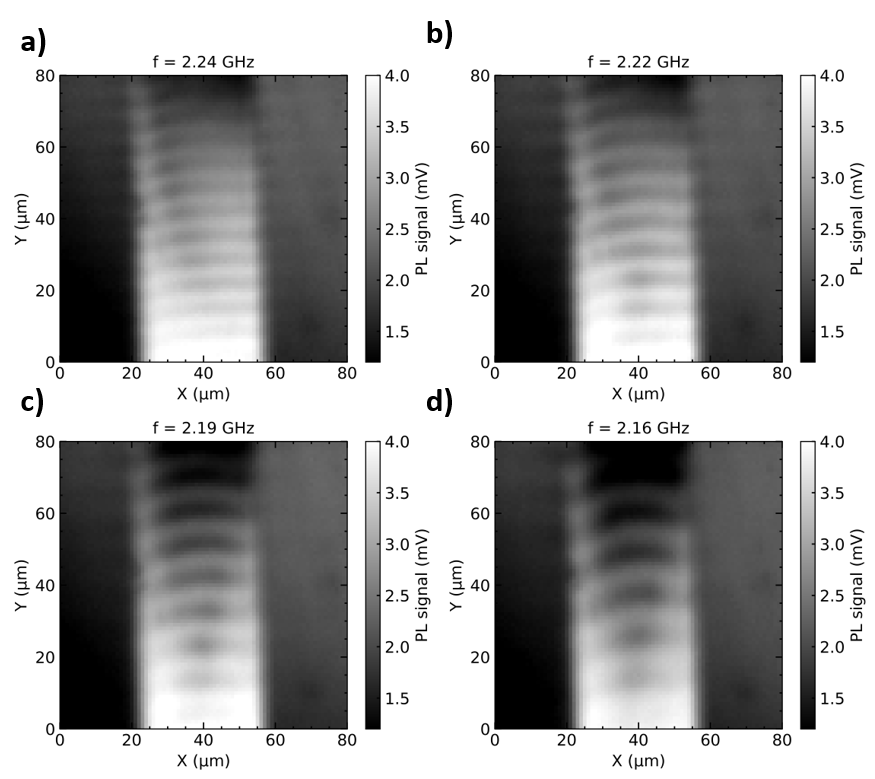}
    \caption{Spatial PL signal measured above the YIG film, when spinwaves excited by the frequency $f$, depending on the external applied field strength, travel through the YIG thin film. }
    \label{fig:A_NV_spin_waves}
\end{figure}

\begin{figure}[htb!]
    \centering
    \includegraphics[width=0.4\textwidth]{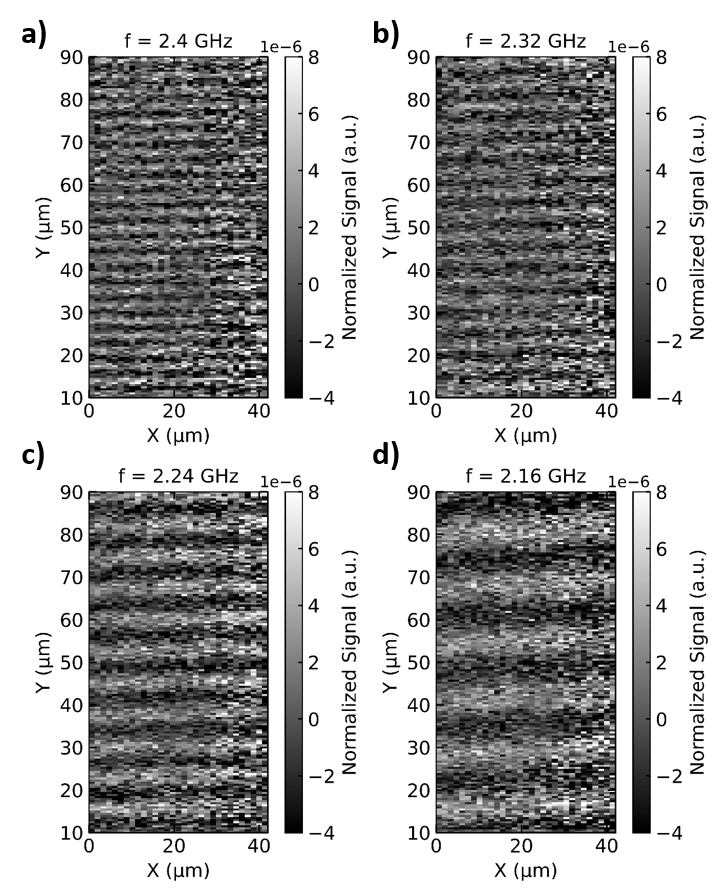}
    \caption{Spatial TR-MOKE signal measured above the YIG film, when spinwaves excited by the frequency $f$, depending on the external applied field strength, travel through the YIG thin film. }
    \label{fig:A_Kerr_spin_waves}
\end{figure}

Figures \ref{fig:A_NV_spin_waves} and \ref{fig:A_Kerr_spin_waves} provide additional images of the observed spin wave wavefronts, captured by the NV setup and the TR-MOKE setup respectively, under various external fields resulting in different excitation frequencies and wavelengths.

\section{TR-MOKE dispersion for fields applied in DE configuration and along the NV axis}
\label{A:Moke DE vs NV}
\begin{figure}[htb!]
    \centering
    \includegraphics[width=0.25\textwidth]{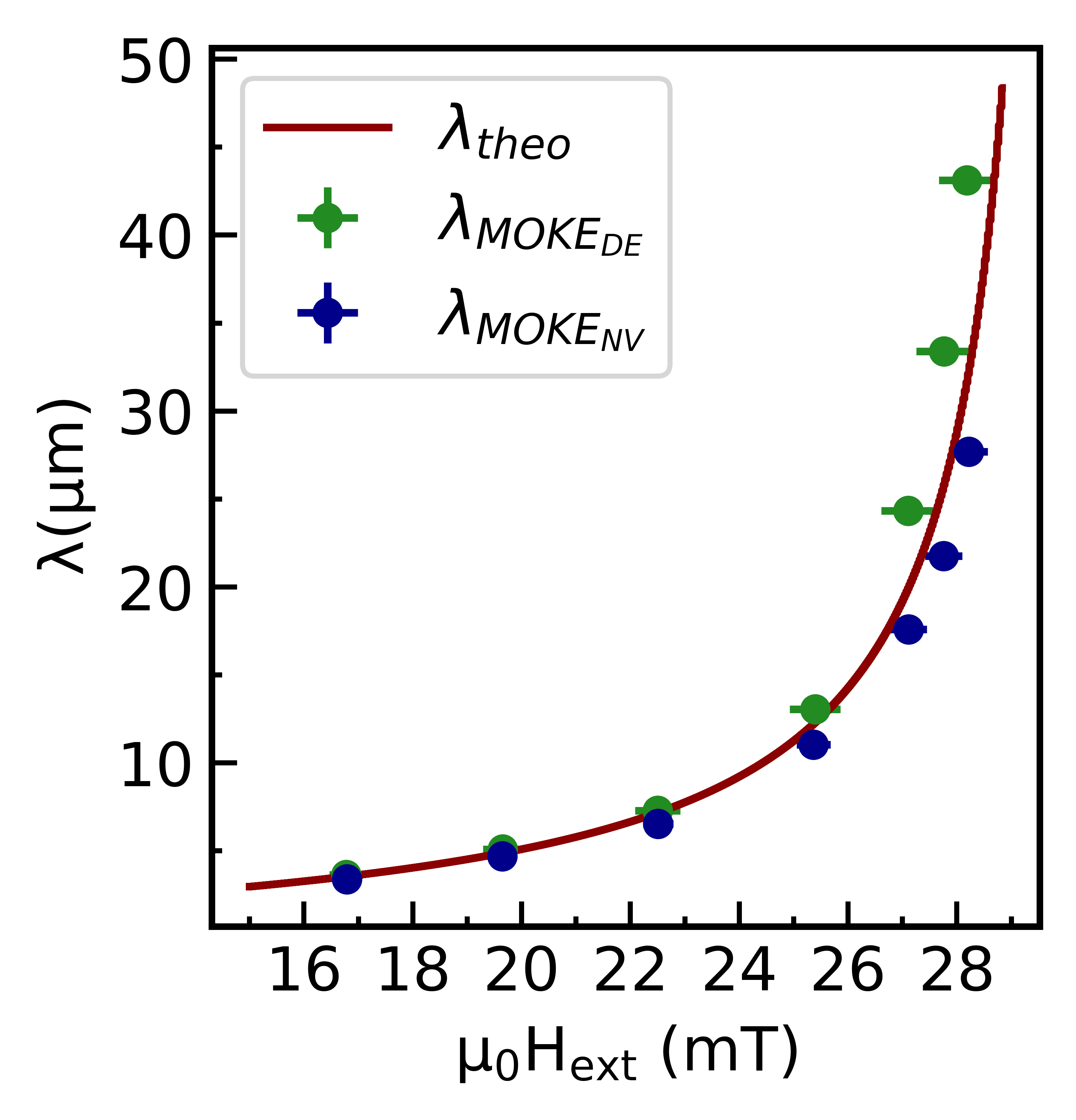}
    \caption{Dependence of the wavelength of the excited spin waves if the external magnetic field is applied in DE configuration (green dots) and along S1 at an angle of $\phi = 35^\circ$ relative to the sample plane. The theoretical dispersion curve (red line) is in good agreement with the fitted wavelengths.}
    \label{fig:moke_comparison}
\end{figure}

In Fig.~\ref{fig:moke_comparison}, the spin wave dispersion extracted for the case where the external magnetic field is applied in DE configuration (green dots) and along S1 at an angle of $\phi = 35^\circ$ relative to the sample plane, aligning it with one of the four possible NV center axes direction (blue dots), is plotted against the theory curve (red line). 
The deviation in the fitted wavelengths arises from errors in the magnetic field calibration. However, within the bounds of this error, the measurements closely match the expected theoretical curve, demonstrating the negligible effect of the perpendicular component of the external field with respect to the thin film plane.

\section{Fitting method}
\label{A:fit model}
The wavefronts $A(y)$ of the NV- and TR-MOKE measurements were both fitted by 
\begin{equation}
    A(y) = A e^{y-y_0} \mathrm{sin}\bigg(\frac{2\pi}{\lambda}y + \phi_0\bigg) + a_\mathrm{off} + y a_0,
\end{equation}
where $A$ is the wavefront amplitude, $y_0$ the distance of the X-axis from S1, $\lambda$ the wavelength of the spin wave, $\phi_0$ a phase constant, $a_\mathrm{off}$ a constant for the signal offset, and $a_0$ the slope of the signal.

\end{document}